\documentclass[english,twocolumn]{emulateapj}
\usepackage{color}
\usepackage{epstopdf}
\usepackage{amsmath}
\makeatletter
\newcommand\lsim{\mathrel{\rlap{\lower4pt\hbox{\hskip1pt$\sim$}}
\raise1pt\hbox{$<$}}}
\newcommand\gsim{\mathrel{\rlap{\lower4pt\hbox{\hskip1pt$\sim$}}
\raise1pt\hbox{$>$}}}
\newcommand\lleq{\mathrel{\rlap{\raise1pt\hbox{\hskip2pt$\diagup$}}
\raise1pt\hbox{$\lll$}}}

\newcommand{\tot}{\mathrm{tot}}
\newcommand{\PN}{1\mathrm{PN}}

\newcommand{\Ham}{\mathcal{H}}

\newcommand{\out}{\rm out}
\newcommand{\inner}{\rm in}
\newcommand{\itot}{i_\textnormal{tot}}

\newcommand{\msun}{M_\odot}

\newcommand{\fme}{\mathit{f}_{me_1}}

\makeatother
\shorttitle{Resonant 1PN Eccentricity Excitation in Hierarchical Triples}
\shortauthors{Naoz et al.}
\makeatother
\begin{document}
\title{Resonant Post-Newtonian Eccentricity Excitation \\ in Hierarchical Three-body Systems}

\author{Smadar Naoz\altaffilmark{1,2,$\dagger$}, Bence Kocsis\altaffilmark{1},  Abraham Loeb\altaffilmark{1},  Nicol\'{a}s Yunes\altaffilmark{3}}
\altaffiltext{1}{Institute for Theory and Computation, Harvard-Smithsonian Center for Astrophysics, 60 Garden St.; Cambridge, MA, USA 02138}
\altaffiltext{2}{Center for Interdisciplinary Exploration and Research in Astrophysics (CIERA), Northwestern University, Evanston, IL 60208, USA}
\altaffiltext{3}{Department of Physics, Montana State University, Bozeman, MT 59718, United States.}
\altaffiltext{$\dagger$}{ Einstein Fellow}
\email{snaoz@cfa.harvard.edu  }
\begin{abstract}

We study the secular, hierarchical three-body problem to first-order in a post-Newtonian expansion of General Relativity. We expand the first-order post-Newtonian Hamiltonian to leading-order in the ratio of the semi-major axis of the two orbits. In addition to the well-known terms that correspond to the GR precession of the inner and outer orbits, we find a new secular post-Newtonian interaction term that can affect the long-term evolution of the triple. We  explore the parameter space for highly inclined and eccentric systems, where the Kozai-Lidov mechanism can produce large-amplitude oscillations in the eccentricities. The standard lore, i.e.,~that General Relativity effects suppress eccentricity, is only consistent with the parts of phase space where the General Relativity timescales are several orders of magnitude shorter than the secular Newtonian one. In other parts of phase space, however, post-Newtonian corrections combined with the three body ones, can excite eccentricities. In particular, for systems where the General Relativity timescale is comparable to the secular Newtonian timescales, the three-body interactions give rise to a resonant-like eccentricity excitation. Furthermore, for triples with a comparable-mass inner binary,
 where the eccentric Kozai-Lidov mechanism is suppressed,  post-Newtonian corrections can further increase the eccentricity and lead to orbital flips even when the timescale of the former is much longer than the timescale of the secular Kozai-Lidov quadrupole perturbations.
\end{abstract}

\section{Introduction}
\label{intro}

Triple stellar systems are believed to be very common in Nature \citep[e.g.,][]{T97,Eggleton+07}. From dynamical stability arguments, these systems must be hierarchical triples, in which the (inner) binary is orbited by a third body on a much wider orbit. Probably, more than 50\% of the bright stars we see are at least (double) binary systems \citep{T97,Eggleton+07}. Given the selection effects against finding faint and distant companions, we can be reasonably confident that the number of triple systems is actually substantially greater than that observed. \citet{T97} showed that $40\%$ of binary stellar systems with period $<10$~days, in which the primary is a dwarf ($0.5 -1.5\,M_{\odot}$), have at least one additional companion. He also found that the fraction of triples and higher multiples among binaries with period ($10-100\,$day) is $\sim10\%$. Moreover, \citet{Pri+06} have surveyed a sample of contact binaries, and noted that 42$\pm5\%$ of 151 of them brighter than 10 mag.~are at least triples. We can then conclude that many close stellar binaries with two compact objects are likely produced through triple evolution.

Long-term stability of triple system requires hierarchical configurations: an ``inner'' binary (with masses $m_1$ and  $m_2$) in a nearly Keplerian orbit with semi-major axis (SMA) $a_1$, and an ``outer'' binary in which $m_3$ orbits the center of mass of the inner binary, with SMA $a_2\gg a_1$. Another stability condition is that the perturber does not make close approaches to the inner binary orbit. In this stability regime, a highly inclined perturber can produce large-amplitude  oscillations in the eccentricity and inclination of the system, the so-called {\emph{Kozai-Lidov mechanism}} \citep{Kozai,Lidov}.

The Kozai-Lidov mechanism is an important example of a secular effect (i.e., coherent interaction on timescales long compared to the orbital period) that is common in hierarchical triple systems but absent from two-body dynamics. This process has been proposed as an important element in the evolution of triple stars \citep[e.g.,][]{Har69,Mazeh+79,1998KEM,Dan,PF09,Tho10,Naoz+11sec,Prodan+12,Sharpee+12} and extrasolar planetary systems with an additional distant stellar
companion \citep[e.g.,][]{Hol+97,Dan,Wu+07,Takeda,Naoz+12bin}. In addition, the Kozai-Lidov mechanism has been suggested to play an important role in both the growth of black holes (BHs) at the centers of dense stellar clusters and the formation of short-period binary BHs \citep{Wen,MH02}.  Furthermore, \citet{Iva+10} showed that the most important formation mechanism for BH X-ray binaries in globular clusters may be triple-induced mass transfer in a BH-white dwarf binary.

Given the hierarchical galaxy formation paradigm, and the strong evidence that a high abundance of the local galaxies host supermassive BHs (SMBHs), one expects that major galaxy mergers should inevitably result in the formation of SMBH binaries or multiples \citep{Valtonen96,Loren,Kulkarni+12,Dotti+12}. \citet{Bla+02} showed that the Kozai-Lidov mechanism plays an important role in the evolution of SMBH triples, where high eccentricity induced by the outer perturber can lead to a more efficient merger rate, due to gravitational wave (GW) emission \citep[see also][]{Seto12}. Also, recently \citet{AP12} showed that secular three-body effects play an important role in the evolution of binary compact objects near SMBH.

GWs emitted during Kozai-Lidov--induced, highly eccentric orbits of compact binaries might be detectable using LIGO\footnote{\url{http://www.ligo.caltech.edu/}} and VIRGO\footnote{\url{http://www.ego-gw.it/}} (e.g., \citealt{Wen} but see \citealt{Ilya08} and \citealt{OLeary_06}), pulsar timing arrays \citep[e.g.,][]{Finn+10,Amaro-Seoane+11,2012ApJ...752...67K}, and  future space-based GW observatories, such as eLISA/NGO  \citep{eLISA1,eLISA2}\footnote{\url{http://elisa-ngo.org/}}. In fact, GWs associated with eccentric orbits are stronger and have a very different spectrum relative to their circular counterparts for sources at the same distance and with the same mass and spin. This may allow for the GW detection of eccentric inspirals with higher masses, larger SMAs or farther away from Earth relative to their quasi-circular counterparts \citep{Arun:2007qv,Arun:2007hu,Yunes:2009yz,2009MNRAS.395.2127O,2011arXiv1109.4170K}. Using GW information emitted by the close binary, it might be possible to constrain the parameters of the third body, such as its mass or distance, provided the GW signal-to-noise ratio is sufficiently high \citep{Yunes:2010sm,Galaviz+11}.

The Kozai-Lidov mechanism is therefore tremendously important and there is still much to be understood. Recently, \citet{Naoz11,Naoz+11sec} showed that an eccentric outer orbit (and even a circular one with comparable mass inner binary)
can behave significantly differently than previously assumed, the so-called ``eccentric Kozai-Lidov mechanism".  Specifically, they showed that the inner orbits can flip from prograde to retrograde and back, and can also reach extremely high eccentricities close to unity, and the system behaves  chaotically  \citep{LN}.
Most previous secular three body dynamics studies that incorporated GR effects did so through a pseudoÐpotential, constructed mainly to model accretion disks and 1st post-Newtonian  ($\PN$) shifts in the innermost stable circular orbit \citep{NW91,Artemova+96,MH02}. It has been shown that the 1PN precession of the inner body may play an important role in secular evolution \citep[e.g.][]{Ford00,MH02,Bla+02,Mardling07,Dan,Zhang+13}. Here we expand our investigation to include both the eccentric Kozai-Lidov mechanism and the three body  $\PN$ effects. We show here (\S \ref{sec:evol} and Appendix \ref{sec:2body}) that although this pseudo--potential does capture some $\PN$ effects, such as the precession rate, the full $\PN$ three-body Hamiltonian introduces other corrections that cannot be modeled with this potential.

In this paper, we study the consistent inclusion of $\PN$ terms in the secular dynamical evolution of hierarchical triple systems. We restrict attention to the $\PN$ approximation of the three-body Hamiltonian. While it is well established that the eccentricity and inclination are constant in the $\PN$ two-body problem \citep{DD85}, it is not true for hierarchical triples. In addition to the standard GR precession of the inner and outer orbits, the $\PN$ corrections lead to a new secular interaction between the inner and outer binaries that affects their long-term evolution. We find that  the standard lore, i.e.,~that GR effects suppress eccentricity, is only true when the GR timescales are several orders of magnitude shorter than the secular Newtonian ones. When the GR timescales are comparable to the secular Newtonian ones, we show that three-body interactions generally give rise to a resonant-like eccentricity excitation  \citep[see also][]{Ford00}. We will be using the term ``resonance" here to describe the rapid excitations of the inner orbit's eccentricity, which occurs when the $\PN$ timescales are comparable to the secular Newtonian timescales.
We demonstrate that even for systems with comparable inner binary masses,
where the Kozai-Lidov mechanism is suppressed, and even when the GR timescales are much longer than the secular Newtonian ones, $\PN$ corrections continue to excite the eccentricity.

This paper is organized as follows. We begin with a definition of the parameters used to describe a hierarchical triple system based on Newtonian and $\PN$ three-body Hamiltonians (\S \ref{Sec:Ham}). We then show that three-body evolution is modified by $\PN$ effects (\S \ref{sec:evol}). We discuss the different time-scales corresponding to the $\PN$ effects, and identify the region in phase space where important deviations might arise due to these terms (\S \ref{sec:times}). We then show that $\PN$ terms can, in many cases, excite the eccentricity of the inner orbit instead of suppressing it (\S \ref{sec:eexcit}). We conclude with a discussion in \S \ref{sec:dis}.

\section{Hamiltonian Perturbation Theory for Hierarchical Triple Systems}\label{Sec:Ham}

\begin{figure}[htb]
\begin{center}
\includegraphics[width=8.5cm,clip=true]{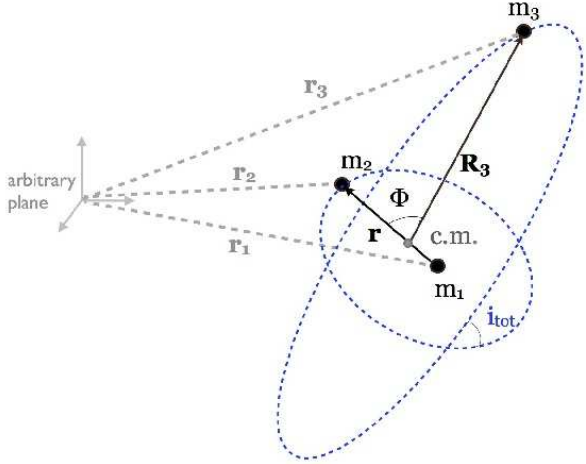}
\caption{Coordinate system used to describe the hierarchical triple system ({\it not to scale}). Here 'c.m.' denotes the center of mass of the inner binary, containing objects of masses $m_1$ and $m_2$. The separation vector $\bf{r}$ points from $m_1$ to $m_2$; $\bf{R}_3$ points from 'c.m.' to $m_3$. The angle between the vectors ${\bf r}$ and ${\bf R}_3$ is $\Phi$. The distances from the bodies to a field point are labeled by $\bf{r}_{1}$, $\bf{r}_{2}$ and $\bf{r}_{3}$.} \label{fig:config}
\end{center}
\end{figure}

A triple system consists of a binary (with masses $m_1$ and $m_2$) and a third body (with mass $m_3$) in orbit about the center of mass of the former. It is convenient to describe the orbits using Jacobi coordinates \citep[][]{MD00}. Let ${\bf r}$ be the relative position vector from $m_1$ to $m_2$ and ${\bf R}_3$ be the position vector of $m_3$ relative to the center of mass of the inner binary \citep[see for more details ][]{Naoz+11sec}, as shown in Fig.~\ref{fig:config}.

In the PN approximation, corrections to Newtonian mechanics arise in powers of $(v/c)^n$, where $v$ is the orbital velocity and $c$ is the speed of light, with $n\geq2$ an integer. Here we concentrate on the $\PN$ order corrections to Newtonian motion, which are  ${\cal{O}}(v^{2}/c^2)$  relatively smaller than the  Newtonian terms. The Hamiltonian can then be divided into a Newtonian part ($\Ham_{\rm N}$) and a $\PN$ part ($\Ham_{\PN}$):
\begin{equation}
\Ham_{\tot,\PN}=\Ham_{\rm N}+\Ham_{\PN} \,,
\label{tot-Ham}
\end{equation}
where the Newtonian part is simply
\begin{equation}
\Ham_{\rm N}= \frac{1}{2}\sum_{\i=1}^{3}\frac{p_i^2}{m_i}-\frac{1}{2}\sum_{i,j\neq i}^{3}\frac{k^2 m_i m_j}{r_{i j}} \ ,
\end{equation}
and the $\PN$ part is \citep[e.g.][]{Schafer87,Moore93,Lousto+08}:
\begin{eqnarray}\label{eq:1PN3b}
\Ham_{\PN}&= &-\frac{1}{8 c^2}\sum_{i=1}^{3}m_i\left(\frac{p_i^2}{m^2_i}\right)^2  -\frac{k^2}{4c^2}\sum_{i,i\neq j} \frac{m_i m_j}{r_{ij}} \bigg\{ 6\frac{p^2_i}{m_i^2}   \nonumber  \\
&-& 7 \frac{({\bf{p}}_i \cdot {\bf{p}}_j)}{m_i m_j}-\frac{ ({\bf{n}}_{ij} \cdot {\bf{p}}_i )({\bf{n}}_{ij} \cdot {\bf{p}}_j )  }{m_i m_j} \bigg \}  \nonumber \\
&+&\frac{k^4}{2c^2}\sum_{i,j\neq i, k'\neq i} \frac{m_i m_j m_{k'}}{r_{ij}r_{ik'}}   \,.
\end{eqnarray}
In these equations, $k^2$ is the gravitational constant, $\bf{r}_{i j}$ ($r_{ij}$) is the relative position vector (magnitude) from mass $m_i$ to $m_j$, ${\bf p}_i$ (${p}_i$) is the momentum vector (magnitude) of mass $m_i$ in an arbitrary plane (we shall later transform to center-of-mass coordinates). In the $\PN$ Hamiltonian, $i,j$ and $k'$ run from $1$ to $3$ (the three masses), where $k'$ is an index while $k^2$ is the gravitational constant, and ${\bf n}_{ij}={\bf{r}}_{ij}/r_{ij}$.

Many gravitational triple systems are in a hierarchical configuration: two objects orbit each other in a relatively tight inner binary while the third object is on a much wider orbit. If the third object is sufficiently distant, an analytic, perturbative approach can be used to calculate the evolution of the system over long timescales (relative to the orbital period).  In the usual secular approximation \citep[e.g.,][]{Maechal90}, the three orbiting objects torque each other and exchange angular momentum, but not energy.  Therefore, on timescales much longer than their orbital periods, the eccentricity and orientation can change, but not the SMA.

Given this, the orbital motion of a triple system can be divided into two separate Keplerian orbits: the relative orbit of bodies~1 and~2, and the orbit of body~3 around the center of mass of the  system.
 The Hamiltonian for the system can then be decomposed accordingly into two Keplerian Hamiltonians plus a coupling term that describes the (weak) interaction between the two orbits. Let the SMAs of the inner and outer orbits be $a_1$ and $a_2$, respectively. Then, the coupling term in the Hamiltonian can be written as a power series in the ratio of the SMAs $\alpha=a_1/a_2$ \citep[e.g.,][]{Har68}. In a hierarchical system, by definition, this parameter $\alpha$ is small.

The Newtonian part of the Hamiltonian, expanded in powers of $\alpha$, is
\citep[e.g.,][]{Har68},
\begin{eqnarray}
\label{eq:Ham}
\Ham_{\rm N}&=&-\frac{k^2 m_1m_2}{2a_1}-\frac{k^2 m_3(m_1+m_2)}{2a_2} \\ \nonumber
&& - \frac{k^2}{a_2} \sum_{j=2}^\infty\alpha^j M_j \left(\frac{r}{a_1}\right)^j\left(\frac{a_2}{R_3}\right)^{j+1}P_j(\cos{\Phi}) \ ,
\end{eqnarray}
where  $P_j$ are Legendre polynomials, $\Phi$ is the angle between ${\bf r}_1$ and ${\bf r}_2$ (see Fig.~\ref{fig:config}) and
\begin{equation}
\label{eq:Mj}
M_j=m_1m_2m_3\frac{m_1^{j-1}-(-m_2)^{j-1}}{(m_1+m_2)^{j}} \ .
\end{equation}
Note that most secular studies follow the convention of \citet{Har69} and choose the Hamiltonian to be the negative of the total energy, so that $\Ham > 0$ for bound systems. Here we did {\it not} follow this convention. The equations of motion  in \citet{Naoz+11sec} did use this convention, and thus, a reader that wishes to combine the two sets of equations need to introduce a minus sign to one of the sets.

We adopt canonical variables, known as Delaunay's elements, which provide a particularly convenient dynamical description of hierarchical three-body systems \citep[e.g.][]{3book}. The coordinates are chosen to be the mean anomalies, $l_1$ and $l_2$, the arguments of periastron, $g_1$ and $g_2$, and the longitudes of ascending nodes, $h_1$ and $h_2$, where subscripts $1,\,2$ denote the inner and outer orbits, respectively.
Their Newtonian conjugate momenta are
\begin{eqnarray}
L_1&=&\frac{m_1 m_2}{m_1+m_2}\sqrt{k^2 (m_1+m_2)a_1} \ , \label{eq:L1}\\ \nonumber
L_2&=&\frac{m_3(m_1+ m_2)}{m_1+m_2+m_3}\sqrt{k^2 (m_1+m_2+m_3)a_2} \ , \label{eq:L2}
\end{eqnarray}
\begin{equation}\label{eq:G12}
G_1=L_1\sqrt{1-e_1^2} \ , \quad G_2=L_2\sqrt{1-e_2^2} \ ,
\end{equation}
and
\begin{equation}
H_1=G_1\cos{i_1} \ , \quad H_2=G_2\cos{i_2} \  ,
\end{equation}
respectively, where $e_1$ ($e_2$) is the inner (outer) orbital eccentricity. Note that $G_1$ and $G_2$ are also the magnitudes of the angular momentum vectors (${\bf G}_1$ and ${\bf G}_2$), and $H_1$ and $H_2$ are the $z$-components of these vectors. The following geometric relations between the momenta follow from the law of cosines:
\begin{eqnarray}
\cos{\itot}&=&\frac{G_\tot^2-G_1^2-G_2^2}{2G_1G_2} \ , \label{eq:cosi} \\
H_1&=&\frac{G_\tot^2+G_1^2-G_2^2}{2G_\tot} \ , \label{eq:H1} \\
H_2&=&\frac{G_\tot^2+G_2^2-G_1^2}{2G_\tot} \ , \label{eq:H2}
\end{eqnarray}
where ${\bf G}_{\tot}={\bf G}_1+{\bf G}_2$ is the (conserved) total angular momentum vector, and the angle between $\bf{G}_1$ and $\bf{G}_2$ defines the mutual inclination $i_\tot=i_1+i_2$.  From Eqs.~(\ref{eq:H1}) and~(\ref{eq:H2}) we find that the inclinations $i_1$ and $i_2$ are determined by the orbital angular momenta via
\begin{eqnarray}
\cos i_1&=&\frac{G_{\tot}^2+G_1^2-G_2^2}{2G_{\tot}G_1} \ , \label{eq:i1} \\
\cos i_2&=&\frac{G_{\tot}^2+G_2^2-G_1^2}{2G_{\tot}G_2} \ . \label{eq:i2}
\end{eqnarray}
In addition to these geometrical relations we also have that
\begin{equation}
\label{eq:z-sum-to-total}
H_1 + H_2 = G_{\tot} = {\rm const} \,
\end{equation}
since we are here neglecting dissipative effects such as GW radiation-reaction, and thus, the Hamiltonian is conserved. Given this parameterization, the Hamiltonian or canonical equations describe the orbital motion via
\begin{eqnarray}
\label{eq:Canoni}
\frac{dL_j}{dt}=-\frac{\partial \Ham}{\partial l_j} \ , \quad \frac{dl_j}{dt}=\frac{\partial \Ham}{\partial L_j} \ , \\
\frac{dG_j}{dt}=-\frac{\partial \Ham}{\partial g_j} \ , \quad \frac{dg_j}{dt}=\frac{\partial \Ham}{\partial G_j} \ , \\
\frac{dH_j}{dt}=-\frac{\partial \Ham}{\partial h_j} \ , \quad \frac{dh_j}{dt}=\frac{\partial \Ham}{\partial H_j} \ ,
\label{eq:Canoni3}
\end{eqnarray}
where $j=1,2$.

The secular Hamiltonian (both the Newtonian and the $\PN$ parts) is given by taking Eq.~\eqref{tot-Ham} expanded in powers of $\alpha$ and averaging over the rapidly varying $l_1$ and $l_2$. The averaging technique we use is known as the {\emph{Von Zeipel transformation}} \citep[for more details, see][]{bro59,Naoz+11sec}, also see Appendix \ref{sec:VZ}, a canonical transformation that eliminates the rapidly-oscillating parts of $\Ham$. We apply this transformation twice, leading to a Hamiltonian that is the double average of the original Hamiltonian over both orbital periods. We thus refer to the resulting quantity as the ``double-averaged Hamiltonian.''

The double-averaged Newtonian Hamiltonian [Eq.~(\ref{eq:Ham})], up to octupole order \citep[i.e.,~up to ${\cal{O}}(\alpha^3)$ beyond the leading order term proportional to $a_1^{-1}$, see][]{Naoz+11sec}\footnote{Note that  \citet{KM99} showed that the von Zeipel transformation results in  higher orders terms proportional to  $\alpha^{7/2}$, however here we consider only $\mathcal{O}(\alpha^3)$ level of perturbations. }
 can be written as:
\begin{equation}
\label{eq:HamN-ave}
\bar{\Ham}_{\rm N}=\bar{\Ham}_{\rm quad}^{\rm N}+\bar{\Ham}_{\rm oct}^{\rm N}
\end{equation}
where
\begin{eqnarray}\label{eq:Hquad}
 \bar{\Ham}^{\rm N}_{\rm quad}& =&- C_2 \{ \left( 2 + 3 e_1^2 \right) \left( 3 \cos^2 i_\tot - 1 \right) \nonumber \\
 & +& 15 e_1^2 \sin^2 i_\tot \cos(2 g_1) \} \ , \\
  \bar{\Ham}^{\rm N}_{\rm oct}&=& \frac{15}{4}\epsilon_M e_1 C_2  \{ A \cos \phi + 10 \cos i_\tot \sin^2 i_\tot \nonumber \\
   &\times& (1-e_1^2) \sin g_1 \sin g_2 \} \ .
\end{eqnarray}
Note that Equation (\ref{eq:Hquad}) has a minus sign compare to \citet{Naoz+11sec} that used the sign convention for which the Hamiltonian is positive.
Here, we did not include the terms which correspond to the Keplerian orbital energy of the three objects which depend on only the SMAs, and are constant in the secular approximation without dissipative effects.
Furthermore, we have defined
\begin{eqnarray}
\label{eq:epsiM}
\epsilon_M &=& \left(\frac{m_1-m_2}{m_1+m_2}\right)\left( \frac{a_1}{a_2}\right)\frac{e_2}{1-e_2^2} \ ,\\
\label{eq:C2}
 C_2&=&\frac{k^4}{16}\frac{(m_1+m_2)^7}{(m_1+m_2+m_3)^3}\frac{m_3^7}{(m_1 m_2)^3}\frac{L_1^4}{L_2^3 G_2^3} \ , \\
A&=& 4+3e_1^2-\frac{5}{2}B\sin i_\tot^2 \ , \\
B&=&2+5e^2_1-7e_1^2\cos(2g_1) \ ,
\end{eqnarray}
and
\begin{equation}
\cos \phi=-\cos g_1\cos g_2 -\cos i_\tot \sin g_1 \sin g_2 \ .
\end{equation}
Note that the octupole coefficient in \citet{Ford00},  is simply $C_3=C_2(\epsilon_M /e_2)15/4$.
Also, following our definitions (see Figure \ref{fig:config}) $m_{1}$ and $m_{2}$ refer to the component masses of the \emph{inner} orbit, while $e_{2}$ refers to the eccentricity of the \emph{outer} orbit.
In the test-particle limit (i.e., $m_1\gg m_2$) $\epsilon_M$ [Eq.~(\ref{eq:epsiM})] reduces to the octupole coefficient introduced in \citet{LN} and \citet{Boaz2},
\begin{equation}
\epsilon=\left(\frac{a_1}{a_2}\right)\frac{e_2}{1-e_2^2} \ .
\end{equation}
In these Hamiltonians (and in the following $\PN$ parts), we have eliminated the nodes (i.e., $h_1$ and $h_2$) by using the conservation of total angular momentum, which leads to $h_1-h_2=\pi$. As shown in \citet{Naoz+11sec} this can be done only as long as one does not conclude that the conjugate momenta are constant \citep[e.g.,][]{Dirac}. The full equations of motion up to the Newtonian octupole order are presented in \citet{Naoz+11sec}.

The averaged $\PN$ Hamiltonian can be separated into different terms. First, let us use the fact that for Keplerian orbits the momentum can be related to the radius and SMA; for the inner orbit, we can write $p_{\inner}=\mu_{\inner} \sqrt{k^2 (m_1+m_2)(2/r-1/a_1)}$, where $\mu_{\inner}$ is the reduced mass of the inner orbit and a similar relation can be written to the outer orbit. Second, we substitute this relation into the three body $\PN$ Hamiltonian, i.e.,  Eq.~(\ref{eq:1PN3b}). After transforming to the center of mass frame, the $\PN$ corrections is expanded in powers of $\alpha$ up to relative ${\cal{O}}(\alpha^{3})$. This produces a similar expansion to Eq.~(\ref{eq:Ham}) for the $\PN$ Hamiltonian, but due to its length we have chosen not to present it here. To investigate the long-term dynamics of the three-body system, we eliminate all terms with short-periods in the Hamiltonian, which depend on the rapidly changing  $l_1$ and $l_2$, using a double Von Zeipel transformation \citep{bro59}, see for more details Appendix \ref{sec:VZ}. In doing so, we must first calculate the angle between the vectors ${\bf p}_{\inner}\cdot {\bf p}_{\out}$ and ${\bf p}_{\out}\cdot {\bf r}$ and ${\bf p}_{\inner}\cdot {\bf R}_3$, where ${\bf p}_{\inner}$ (${\bf p}_{\out}$) is the momentum of the inner (outer) orbit, as defined in the invariable plane.

The leading-order term in an $\alpha \ll 1$ expansion is proportional to $a_1^{-2}$ in the double-averaged $\PN$ Hamiltonian. Keeping all terms up to $\mathcal{O}(\alpha^3)$ beyond leading gives
\begin{equation}
\bar\Ham^{\PN} = \bar{\Ham}_{a_1^{-2}}^{\PN} + \bar\Ham_{a_{1} a_2}^{\PN} + \bar\Ham_{a_2^{-2}}^{\PN}  \bar + \Ham_{\rm int}^{\PN}
\end{equation}
where
\begin{eqnarray}\label{eq:1PNa1}
\bar{\Ham}_{a_1^{-2}}^{\PN} &=&\frac{k^4 \mu_{\inner} \left(15 {m_1}^2+29 {m_1} {m_2}+15 {m_2}^2\right)}{8 {a_1}^2 c^2 } \\
&-&\frac{3k^4 {m_1} {m_2} ({m_1}+{m_2})}{{a_1}^2 c^2 \sqrt{1-{e_1}^2}} \ , \nonumber \\
\bar{\Ham}_{a_1a_2}^{\PN}&=&\frac{k^4  {m_1} {m_2} {m_3} (2 ({m_1}+{m_2})+3 {m_3})}{4 {a_1} {a_2} c^2 ({m_1}+{m_2}+{m_3})} \ , \\\label{eq:1PNa2}
\bar\Ham_{a_2^{-2}}^{\PN} & =& \frac{k^4
\mu_{\out} \left(15 ({m_1}+{m_2})^2+29 ({m_1}+{m_2}) {m_3}+15 {m_3}^2\right)}{8 {a_2}^2 c^2 ({m_1}+{m_2}+{m_3})} \nonumber \\
& -& \frac{3 k^4 ({m_1}+{m_2}) {m_3} (m_1+m_2+m_3)}{{a_2}^2 c^2 \sqrt{1-{e_2}^2}} \,  \\
\label{eq:1PNint}
\bar \Ham_{\rm int}^{\PN} &=& \frac{k^2}{4 a_2^3 c^2  \left(1-{e_2}^2\right)^{3/2} (m_1+m_2)} \bigg\{ {G_1} {G_2} [8 ({m_1}+{m_2}) \nonumber \\
&+&6 {m_3}] \cos i_{\tot} +\frac{a_1 k^2 m_1 {m_2} {m_3}  } {8 ({m_1}+{m_2})} ({\fme}-3 {\fme} \cos^2 i_\tot   \nonumber \\
&+&9 {e_1}^2 \left({m^2_1}+m_1m_2+{m^2_2}) {\cos} (2 {g_1}) \sin^2 i_\tot \right)\bigg\} \ ,
\end{eqnarray}
and where
\begin{eqnarray}
\mu_{\inner}&=&\frac{m_1m_2}{m_1+m_2} \ ,
 \\
\mu_{\out}&=&\frac{m_3(m_1+m_2)}{m_1+m_2+m_3} \ ,
 \\
\fme&=&(2-5 e_1^2) (m_1^2+m_2^2)-3 (2-e_1^2) m_1m_2
\end{eqnarray}
Here $\bar \Ham_{\rm int}^{\PN}$ includes all terms of $\mathcal{O}(\alpha^{5/2})$ and $\mathcal{O}(\alpha^3)$ beyond $\bar{\Ham}_{a_1^{-2}}^{\PN}$, since  $G_1\propto a_1^{1/2}$ and $G_2\propto a_2^{1/2}$ according to Eqs.~(\ref{eq:L1}) and (\ref{eq:G12}).

Not all of the different $\PN$ Hamiltonian terms affect the dynamical evolution of the triple. The $\bar{\Ham}_{a_1a_2}^{\PN}$ term only depends on the masses and the SMAs, i.e.,~it does not depend on the canonical coordinates, and thus, it does not affect the canonical equations, although it does change the total energy of the system. On the other hand, $\bar\Ham_{a_1^{-2}}^{\PN}$ and $\bar\Ham_{a_2^{-2}}^{\PN}$ do contribute to the dynamical evolution, as they clearly depend on $e_{1}$ and $e_{2}$. A possible, (intuitive) physical explanation for this is the following. In the $a_{2} \to \infty$ limit, one would expect only two physical effects: precession of the inner orbit and precession of the outer orbit about the inner binary. These two physical effects arise because of  $\bar\Ham_{a_1^{-2}}^{\PN}$ and $\bar\Ham_{a_2^{-2}}^{\PN}$, and thus, $\bar{\Ham}_{a_1a_2}^{\PN}$, (which satisfies $\bar{\Ham}_{a_1a_2}^{\PN}\gg \bar\Ham_{a_2^{-2}}^{\PN}$ for large $a_2/a_1$) cannot contribute to the motion.

The quantity $\bar\Ham^{\PN}_{\rm int}$ is an ``interaction term,'' in that it represents the coupling between the outer and the inner orbits. Notice that this term would not be present if we had truncated the $\alpha \ll 1$ expansion at ${\cal{O}}(\alpha^{2})$. Notice also that the interaction term does not depend on the argument of periapsis of the outer orbit, $g_2$, just like the quadrupole Newtonian Hamiltonian. Therefore, at quadrupole order, e.g.~for a circular outer perturber, the absence of $g_2$ in the Hamiltonian implies that the outer orbital angular momentum, $G_2$, is conserved \citep[the so-called ``happy coincidence'' of][]{Lidov+76}.

\section{Triple body evolution \\ in post-Newtonian theory}\label{sec:evol}

The secular evolution of a three-body hierarchical system to Newtonian, octupole order was studied in \citet{Naoz+11sec}.
As mentioned in \S \ref{intro}, they showed that the commonly assumed  conservation of the $z$-component of the angular momenta of the inner and outer orbits ($H_1$ and $H_2$)
is only correct in the test-particle approximation to quadrupole order.
Newtonian octupole terms further modulate the eccentricity and inclination oscillations. Specifically, for an eccentric and inclined outer perturber, these terms can lead to extremely high eccentricities and flip the inner orbit from prograde to retrograde. This type of behavior also appears in the test-particle limit for an eccentric orbit \citep[e.g.][]{LN,Boaz2,Naoz+11sec}.

\begin{figure}[htb]
\begin{center}
\includegraphics[width=8.5cm,clip=true]{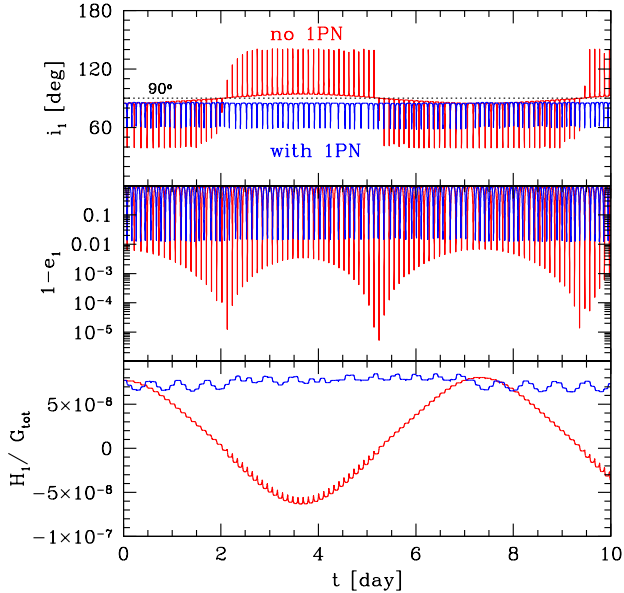}
\caption{An example of the evolution of a triple to Newtonian octupole order neglecting PN contributions (red lines) and including them up to the $\PN$ interactions terms in the double-averaged Hamiltonian (blue lines). The system has an inner binary of $m_1=1$~M$_\odot$,  $m_2=0.001$~M$_\odot$, and a the third object mass is: $m_3=10^4$~M$_\odot$. We set $a_1/R^g_1=10^4$ and $a_2/R^g_3=202$. We set initially $e_2=0.6$, $e_1=0.01$ $g_1=g_2=0^\circ$ and $i_\tot=85^\circ$. We consider, from top to bottom, the inclination of the inner orbit, $i_1$, the eccentricity of the inner orbit in terms of $1-e_1$, and the z-component of the angular momentum normalized to the total angular momentum. Note that we do not plot $G_2$ since in configuration where $m_2\to 0$, $G_2\to G_{\tot}$, \citep{LN}. In this case, the quadrupole Newtonian terms induce eccentricity--inclination oscillations, modulated by the octupole terms, while the $\PN$ effect suppresses them.
}\label{fig:evolve}
\end{center}
\end{figure}

Figure~\ref{fig:evolve} presents the secular evolution of a three-body hierarchical system to Newtonian, octupole order (red lines). We chose a system with inner binary masses $m_1=1$~M$_\odot$ and $m_2=0.001$~M$_\odot$, and an outer binary companion with mass $m_3=10^4$~M$_\odot$. For this system, we set $a_1/R^g_1=10^4$ and $a_2/R^g_3=202$, where, $R^g_1=k^2(m_1+m_2)/c^2$ and $R^g_3=k^2m_3/c^2$ are the gravitational  radii of the inner and the outer orbits, respectively. We also set initially $e_2=0.6$, $e_1=0.01$, $g_1=g_2=0^\circ$ and $i_\tot=85^\circ$. The Newtonian quadrupole terms induce the ``standard'' eccentricity--inclination oscillations, while octupole terms modulate it. As can be seen in the figure, the modulation does not have a precise periodicity and, in fact, the octupole terms introduce the chaotic aspects to the evolution \citep{LN}. When $\PN$ corrections become significant however, the evolutionary orbital tracks can be significantly different (already at quadrupole order).

Different $\PN$ terms have different effects on the evolutionary orbital tracks, where the perturbations to the equations of motion follow from Eqs.~(\ref{eq:Canoni}--\ref{eq:Canoni3}).
$\bar\Ham_{a_1^{-2}}^{\PN}$ gives rise to the standard GR precession of the argument of periapsis of the inner orbit, while $\bar\Ham_{a_2^{-2}}^{\PN}$ is responsible for the precession of the argument of periapsis of the outer orbit,
\begin{eqnarray}
\label{eq:1PNg1}
\frac{dg_1}{dt}\bigg |_{\PN(a_1^{-2})} &=& \frac{3 k^{3} (m_1 + m_2)^{3/2} }{a_1^{5/2} c^2 (1 - e_1^2)} \ ,  \\
\frac{dg_2}{dt} \bigg |_{\PN(a_2^{-2})}&=& \frac{3 k^{3} (m_1 + m_2+m_3)^{3/2} }{a_2^{5/2} c^2 (1 - e_2^2)} \ . \label{eq:1PNs2g1}
\end{eqnarray}
These contributions can be recovered independently from the individual two-body $\PN$ Hamiltonians of the inner and outer binary (see Appendix \ref{sec:2body} Eq.~\ref{eq:1PN2body}),
or from an effective potential, or directly from the $\PN$ metric  \citep[e.g.,][chapter 25 p.~668--670]{Gravitation}.
Other than this precession, $\bar\Ham_{a_1^{-2}}^{\PN}$ and $\bar\Ham_{a_2^{-2}}^{\PN}$, do not directly affect the other orbital elements.\footnote{Note, however, that the precession indirectly affects the evolution of the other orbital elements through  $\bar\Ham^{\rm N}$, as shown below.} The $\bar\Ham^{\PN}_{a_1a_2}$ term just modifies the total energy and does not modify the long--term dynamical evolution at all, as long as dissipative effects are neglected.

In the standard lore, if the GR precession rate of the inner orbit is faster than the quadrupole secular Newtonian   timescales, the GR effect is presumed to suppress the eccentricity growth \citep[for an $m_{2}$ test particle to quadrupole Newtonian order, see][]{Dan}. In Fig.~\ref{fig:evolve} (blue lines) we show an example where this is indeed the case, even when including all $\PN$  terms (see below).
In this example, eccentricity (and orbital flips) are suppressed by the $\PN$ corrections (the variations shown by the blue lines are shorter than red lines).

\begin{figure}[tb]
\begin{center}
\includegraphics[width=8.5cm,clip=true]{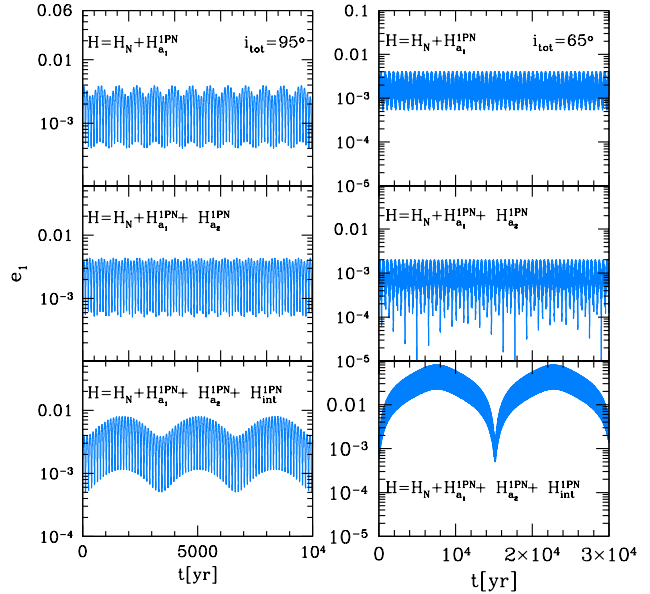}
\caption{Two examples of the time evolution of the system dominated by
$\PN$ effects, the right and left side panels differ only in the
initial relative inclination and the outer orbital separation.
In the top panels we consider the three-body orbit evolution due to
Newtonian dynamics and the lowest--level $\PN$ correction, i.e.,
$\Ham_{N}+\Ham_{a_1^{-2}}^{\PN}$. In the middle panels we add the next
level of approximation $+\Ham_{a_2^{-2}}^{\PN}$, and in the bottom panels we
consider the approximation up to the highest level discussed here,
i.e., $+\Ham^{\PN}_{int}$.
The inner binary contains an object of mass $1\,M_\odot$ and an
object of mass $1\,M_{\rm J}$ (can be considered as a test particle),
while the outer object is a BH with mass of $10^6\,M_\odot$. The inner
orbital separation is $a_1/R^g_1=5.06\times 10^5$, corresponding to
$0.005$~AU. The initial eccentricities are $e_1=0.001$ and $e_2 =0.7$.
The arguments of pericenter of the inner orbit initially set to
$240^\circ$ and outer orbit initially is set to zero.
In the left column, we consider an initial relative inclination of
$i_\tot=95^\circ$, and an outer orbital separation of
$a_2/R^g_3=5.2\times 10^3$, corresponding to $51.4$~AU. In the right
column, we consider an initial relative inclination of $i_\tot=65^\circ$,
and an outer orbital separation of $a_2/R^g_3=4.8\times 10^3$,
corresponding to $47.35$~AU. Observe that as one includes more $\PN$ effects,
qualitatively different behavior emerges.
}
\label{fig:e1_excitGR}
\end{center}
\end{figure}

The usual precession term [Eq.~(\ref{eq:1PNg1})] is not sufficient to model the system, as one must also account for the precession of the outer orbit [Eq.~(\ref{eq:1PNs2g1})] and the other effects introduced by the $\PN$ interaction terms (Appendix \ref{sec:1PNint}). The inclusion of these terms leads to qualitatively different behavior because they directly drive the evolution of inner and outer orbital eccentricity and inclination, while Eq.~(\ref{eq:1PNg1}) [Eq.~(\ref{eq:1PNs2g1})] drives the evolution of only the argument of periapsis of the inner (outer) orbit. Figure~\ref{fig:e1_excitGR} shows the evolution of the eccentricity when different terms in the Hamiltonian are included. We considered a system with parameters $m_1=1\,M_\odot$,  $m_2=1\,M_{\rm J}$ (essentially a test particle), $m_{3} = 10^6\,M_\odot$, $a_1/R^g_1=5.06\times 10^5$  corresponding to $0.005$~AU. In the left column we consider   initial relative inclination $i_\tot=95^\circ$ and a separation of the outer orbit $a_2/R^g_3=5.2\times 10^3$ corresponding to $51.4$~AU. In the right column, we consider initial relative inclination $i_\tot=65^\circ$ and a separation of the outer orbit $a_2/R^g_3=4.8\times 10^3$ corresponding to $47.35$~AU. For the calculation in the two columns, the initial eccentricities are $e_1=0.001$ and $e_2 =0.7$, and the initial argument of pericenter of the inner and outer orbits is set to $240^\circ$ and zero, respectively. This system configuration is such that the $\PN(a_1^{-2})$ timescales for circular orbits ($\sim 59$~yr, for the left column example,  see \S \ref{sec:times} for more details) are shorter than the Newtonian quadrupole ones ($\sim145$~yr  for the left column example). Nevertheless, the secular eccentricity oscillations are still present. The bottom panel shows that the interaction term $\bar \Ham_{\rm int}^{\PN}$ introduces a significant new periodic modulation to the eccentricity evolution. We discuss in more details the conditions in parameter space that lead to this behavior in \S \ref{sec:times} and \ref{sec:eexcit} (note that this system represents the resonance peak of the $95^\circ$ and $65^\circ$ cases of Fig.~\ref{fig:emax} below).

The usual precession term (mostly for the inner) in the presence of three body secular evolution was compared to direct N-body calculation in the literature before \citep[e.g.,][]{Ford00,Zhang+13}. To resolve the effects of the interaction term one needs to be in the the regime where the GR precession time scales are much shorter than the quadrupole precession time scales (see \S \ref{sec:times}). The examples considered in Figure \ref{fig:e1_excitGR} represent $\sim 3 \times 10^8$ of the inner orbital period. Numerical integrations algorithms that conserve the energy over such long timescales in the 3-body post Newtonian regime are not easy to implement or develop, and thus, they are left to future work.

One might worry that a $\PN$ treatment might not be sufficient to model certain regions of phase space, as we have neglected $2$PN and higher PN order terms in the evolution. Such terms become important when the PN perturbation parameter, $v/c$ with $v$ any system velocity, i.e. the pericenter velocities $(v_{p1}/c)^2=(R^g_1+R^g_2)/[a_{1}(1-e_1)]$ and similarly for the outer orbit, are not sufficiently small. In most of our examples we ensured that our systems are well within the PN regime, however, for very eccentric systems, $2$PN and higher PN order terms may be important. In fact, one might naively expect the $\PN$ corrections accounted for here to be negligible if $v/c\ll 1$. This is not so, because although the $\PN$ terms are small at any point in time, their effects may accumulate and become significant over long timescales in the three-body problem.

\section{Timescales}\label{sec:times}

In order to explore the regions of phase space where the $\PN$ effects may be expected to become significant, we compare the various
timescales associated with the individual terms in the Hamiltonian.

\begin{figure*}[htb]
\begin{center}
\includegraphics[width=12cm,clip=true]{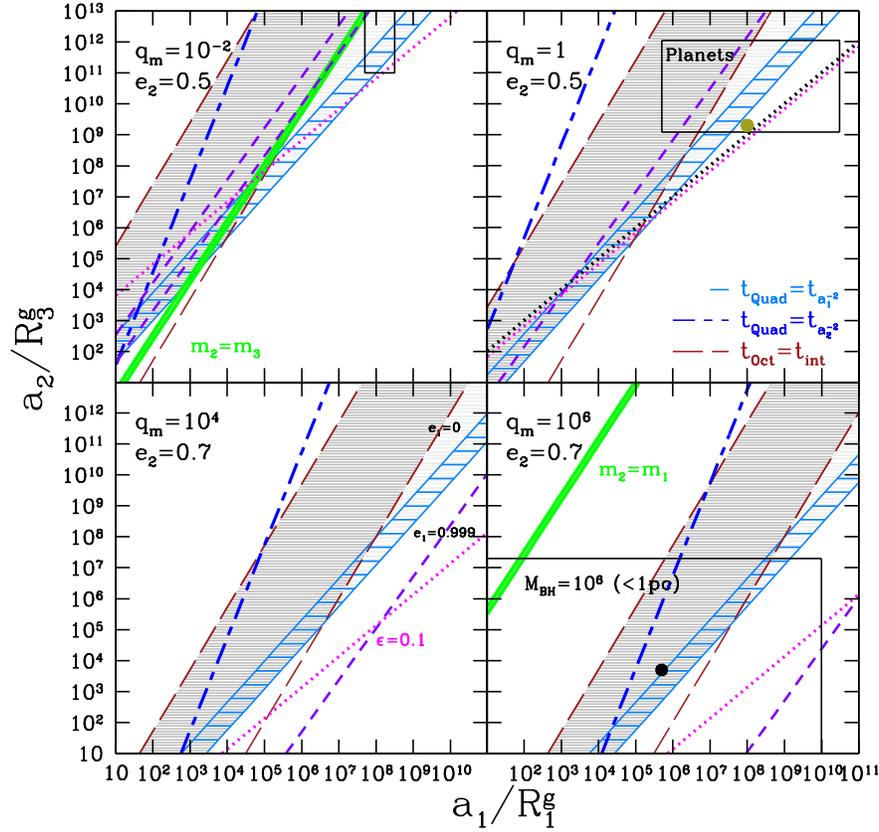}
\caption{
Regions of binary parameters where the various secular Newtonian and $\PN$ effects are expected to become significant. We show the SMAs where the timescales, corresponding to individual terms in the Hamiltonian, are equal to each other for different $a_1/R^g_1$ and $a_2/R^g_3$.
We consider four mass ratios between the outer object and the inner binary $q_m=m_3/m_1=0.01,1,10^4$ and $10^6$ (see labels in each panel) and two different choices for outer orbital eccentricity ($e_2=0.5$, top panels and $e_2=0.7$ lower panels), setting $m_2\to 0$ in all panels. We consider $t^{\rm N}_{\rm quad}=t^{\PN}_{a_1^{-2}}$ (Eq.~(\ref{eq:quada1}), solid blue lines), $t^{\rm N}_{\rm quad}=t^{\PN}_{a_2^{-2}}$ (Eq.~(\ref{eq:quada2}), thick long-short dashed blue lines),  $t^{\rm N}_{\rm oct}=t^{\PN}_{\rm int}$ (Eq.~(\ref{eq:octint}), long dashed brown lines) and $t^{\rm N}_{\rm oct}=t^{\PN}_{a_2^{-2}}$ (Eq.~(\ref{eq:octa2}), short dashed purple lines). For the latter, we show $e_1=0$ (top purple line) in all panels and in the top left panel we also show the eccentric case $e_1=0.999$ (bottom purple line).
The gray and blue shaded areas cover the range $0\leq e_1\leq 0.999$ between the brown shaded lines and the blue solid lines respectively
In the top left panel and bottom right, we also show a green band for which the timescale to shrink the inner orbit by a factor two, due to GW emission, is equal to the quadrupole timescale, which covers the range $0\leq e_1\leq 0.999$.
This line is generated by specifying $m_1=100$~M$_\odot$ and $m_2=m_3=1$~M$_\odot$ (top left panel) and $m_2=m_1=1$~M$_\odot$ (bottom right panel). We also show two stability criteria, $\epsilon=0.1$, dotted magenta lines and the  \citet{Mardling+01} criterion [Eq.~(\ref{eq:Mar}), in the top right panel]. The green dot in the right top panel represents (up to a factor 3) the location in this phase space of the system considered in Fig.~\ref{fig:e1_excite}, while the black dot in the bottom right panel represent the location of the example considered in Fig.~\ref{fig:e1_excitGR}. The rectangle in the top left panel shows the parameter space considered in Figure \ref{fig:e1PSR}. The black rectangles in the top and bottom right panels roughly represent the region in parameter space where planets in stellar binaries (top right) and stars in the galactic nuclei (bottom right) live.
} \label{fig:timescales}
\end{center}
\end{figure*}

The timescale associated with the Newtonian quadrupole term can be estimated from the canonical equations. More precisely, $t^{\rm N}_{\rm quad} \sim 2 \pi G_1/C_{2}$, where $C_{2}$ is given in Eq.~(\ref{eq:C2}):
\begin{equation}\label{eq:tquad}
t^{\rm N}_{\rm quad}\sim \frac{2\pi a_2^3 (1-e_2^2)^{3/2}\sqrt{m_1+m_2}}{ a_1^{3/2} m_3 k}  \ .
\end{equation}

The timescales associated with the Newtonian octupole terms are more difficult to estimate due to their chaotic effect on the orbits. For example, in Fig.~\ref{fig:evolve} the first modulation period is shorter than the second \citep[see,][for more examples]{Naoz+11sec}. However, as demonstrated in \citet{LN}, although the system is chaotic when the octupole terms are included, there are two general features in the evolution: one associated with an octupole timescale, defined below, and a shorter one (see for example their Fig.~7 of surfaces of section, which shows the two different evolutionary behaviors). We define a timescale for the regular part of the Newtonian octupole evolution through the rough estimate $t^{\rm N}_{\rm oct}\sim (4/15) t^{\rm N}_{\rm quad}/\epsilon_M$ for a given inner and outer eccentricity,
\begin{equation}
\label{eq:toct}
t^{\rm N}_{\rm oct}\sim  2\pi \frac{4}{15} \frac{a_2^4 (1-e_2^2)^{5/2} \sqrt{1-e_1^2} (m_1+m_2)^{3/2}}{a_1^{5/2}e_2 k |m_1-m_2| m_3 }\,.
\end{equation}
Note that when the inner binary is very eccentric, these timescales can change by orders of magnitude.
At octupole order, the eccentricity of the outer orbit can also oscillate, although usually these oscillations are small in magnitude. For the remainder of this section, we employ a test-particle approximation for one of the components of the inner binary, i.e.~$m_2\to 0$, for which $e_2={\rm const.}$ \citep{LN}. We will see that this is sufficient to understand the regions of phase space where $\PN$ terms become important.

The $\PN$ timescale can be estimated as in Eq.~\eqref{eq:tquad}, which gives
\begin{eqnarray}
\label{eq:tGRa1}
t^{\PN}_{a_1^{-2}}&\sim&  2\pi \frac{a_1^{5/2} c^2 (1-e_1^2) }{ 3 k^3 (m_1+m_2)^{3/2}} \ , \\
t^{\PN}_{a_2^{-2}}&\sim&  2\pi \frac{a_2^{5/2} c^2 (1-e_2^2) }{ 3 k^3 (m_1+m_2+m_3)^{3/2}} \ , \\
\label{eq:tint}
t^{\PN}_{\rm int} &\sim& \frac{16}{9}\frac{a_2^3 c^2 (1-e_2^2)^{3/2}(m_1+m_2)^{3/2}}{\sqrt{a_1}e_1\sqrt{1-e_1^2} k^3 (m_1^2+m_1m_2+m_2^2)m_3} \ .
\end{eqnarray}
All of these timescales depend on the secularly varying, inner orbital eccentricity, which implies that we need to explore different eccentricity values in phase space. Equations~(\ref{eq:toct}--\ref{eq:tint}) show that the Newtonian and $\PN$ timescales have a simple dependence on the inner and outer orbital eccentricity, on the mass ratio $q_m=m_3/m_1$, and on the SMAs.

If any of the above timescales is much smaller than all others, then the evolution of the triple is dominated by the corresponding term in the Hamiltonian. Next, we examine the three-body parameters where any two timescales are equal, which defines the region where the corresponding two terms are equally important. The corresponding regions are shown in Figure~\ref{fig:timescales} in the $m_2\to 0$ limit.

Equating $t^{\PN}_{a_1^{-2}}$ to $t^{\rm N}_{\rm quad}$ (Eqs.~\ref{eq:tquad} and \ref{eq:tGRa1}) gives a relation between the SMAs which
  normalized to the gravitational radius of the inner and outer binaries,
$R^g_1$ and $R^g_3$, as defined above, can be written as:
\begin{equation}
\label{eq:quada1}
\frac{a_2}{R^g_3}\bigg|_{{\rm quad}=\PN(a_1^{-2})}\sim \left(\frac{1}{3}\right)^{1/3}\left(\frac{a_1}{R^g_1}\right)^{4/3}\frac{1}{q_m^{2/3}}\frac{(1-e_1^2)^{1/3}}{\sqrt{1-e_2^2}} \ .
\end{equation}
This relation is shown by the blue hatched area bounded by solid blue lines in Figure~\ref{fig:timescales} for $0 \leq e_1\leq 0.999$. A resonant-like $\PN$ excitation of eccentricity is possible in this region, as we will show in the next section. For much larger $a_2$ or smaller $a_1$, $t^{\PN}_{a_1^{-2}}\ll t^{\rm N}_{\rm quad}$, and thus the Kozai-Lidov eccentricity excitations are suppressed by the $\PN$ effects.

Next, equating $t^{\PN}_{a_2^{-2}}$ to $t^{\rm N}_{\rm quad}$ gives
\begin{equation}
\label{eq:quada2}
\frac{a_2}{R^g_3}\bigg|_{{\rm quad}=\PN(a_2^{-2})}\sim 3\left(\frac{a_1}{R^g_1}\right)^3\frac{q_m}{(1+q_m)^3}\frac{1}{1-e_2^2} \ .
\end{equation}
This is shown by a blue short--long dashed line in Fig.~\ref{fig:timescales} on the top and bottom panels for $e_2=0.5$ and $0.7$, respectively. For $a_2/R^g_3$ much larger than this value, the Kozai-Lidov oscillations are suppressed and the $\PN$ effects dominate.

Let us next compare $t^{\PN}_{\rm int}$ and $t^{\rm N}_{\rm oct}$ by setting them equal to each other:
\begin{equation}
\label{eq:octint}
\frac{a_2}{R^g_3}\bigg|_{{\rm oct}=\PN(\rm int)}\sim \frac{32}{135\pi}\left(\frac{a_1}{R^g_1}\right)^{2}\frac{1}{q_m}\frac{e_1 e_2}{(1-e_2^2)(1-e_1^2)} \,.
\end{equation}
shown by long-dashed brown lines in  Fig.~\ref{fig:timescales}. The $\PN$ effects are equally important as the Newtonian ones in the gray shaded area in Fig.~\ref{fig:timescales} for $0\leq e_1 \leq 0.999$. This is the regime in which the $\PN$ interaction term introduces qualitatively different behavior in the orbital motion (i.e., modulation) as shown in Fig.~\ref{fig:e1_excitGR}. Outside the gray region in Fig.~\ref{fig:timescales}, the interaction term is negligible. Note that comparing $t^{\PN}_{\rm int}$ with $t^{\rm N}_{\rm quad}$ leads to a vertical line in the phase diagram of Fig.~\ref{fig:timescales}. This is because both timescales have the same dependence on the outer SMA ($\sim a_2^3$), resulting in $a_1/R_1<10$ (not shown). In \S \ref{sec:eexcit}, we explore the parameter space that also leads to excitations in the eccentricity (as shown in Fig.~\ref{fig:e1_excitGR}).

Comparing $t^{\PN}_{a_2^{-2}}$ to the octupole timescale gives usually a longer timescale than the quadrupole
(short--dashed purple lines in Fig.~\ref{fig:timescales}):
\begin{eqnarray}\label{eq:octa2}
\frac{a_2}{R^g_3}\bigg|_{{\rm oct}=\PN(a_2^{-2})}&\sim&\left(\frac{4}{45}\right)^{1/3}\left(\frac{a_1}{R^g_1}\right)^{5/3} \\
&\times& \frac{1}{q_m^{1/3} (1+q_m)^{5/3} }\frac{ e_2^{2/3}}{(1-e_2^2)(1-e_1^2)^{1/3}} \ . \nonumber
\end{eqnarray}
This relation also provides a range of parameters for different values of $e_{1}$ and $e_{2}$, but to avoid cluttering, we plot this timescale only for a  circular inner binary ($e_1=0$).
The boundary shifts to larger $a_2$ in the eccentric case. In top-left panel of Fig.~\ref{fig:timescales} (the $q_m=0.01$ case), we show the range of Eq.~\eqref{eq:octa2} for $e_{2} \in (0, 0.999)$, since, in this case, this ratio is smaller than Eq.~(\ref{eq:octint}).

Figure~\ref{fig:timescales} also shows the regime of validity of the hierarchical triple approximation, where we choose $\epsilon>0.1$ as a rule of thumb for stability (dotted magenta line). For the $q_m=1$ case, this rule of thumb seems to agree with the \citet{Mardling+01} stability criterion, which defines a stable three-body system as one that obeys
\begin{equation}
\label{eq:Mar}
\frac{a_2}{a_1}> 2.8(1+q_m)^{2/5}\frac{(1+e_2)^{2/5}}{(1-e_2)^{6/5}}\left(1-\frac{0.3 i_\tot}{180^\circ}\right ) \ ,
\end{equation}
where in the top right panel we considered this criterion with prograde $i_{\rm tot}=0^\circ$. Note that retrograde motions are usually more stable \citep[e.g.][]{Innanen79,Innanen80,MG12}.

Another consistency requirement for the $\PN$ Kozai-Lidov effects to operate is that gravitational radiation reaction \emph{does not} change the SMA significantly over this timescale. We define $t_{\rm GW1}$,
the GW in-spiral timescale of the inner binary for the SMA to change by factor of two using \citet[][]{Peters64} \citep[see also][]{Arun+09}. Note that $t_{\rm GW1}\to \infty$ when $m_2\to 0$, but the GW inspiral may become very important in the comparable-mass limit and shut off the Kozai-Lidov effect. This is the case above the green bands in the top left and bottom right panels, which show the range of SMAs where $t^{\rm N}_{\rm quad}=t_{\rm GW1}$ for  $m_2=m_3=1$~M$_\odot$ (top left panel) and $m_2=m_1=1$~M$_\odot$ (bottom right panel) for $0\leq e_1\leq 0.999$.
Note that there is a region below the green band where the $\PN$ approximation is insufficient, and where $2$PN and higher PN order corrections need to be included; we leave this to future work.  A similar regime, were the $\PN$ level of approximation is
insufficient, was considered by \citet{Bla+02} for which the evolution was mostly affected by GW emission and resonant eccentricity growth was not observed. 

We conclude that $\PN$ effects may be important for a vast range of parameters as shown in Fig.~\ref{fig:timescales}. Note however, that physical timescales are not shown there; this figure is independent of an overall dimensional scale (e.g. total mass or the scale of the SMAs). The physical timescale may be smaller or larger than the Hubble time depending on the actual system parameters considered.

\section{Excitation of the Inner Orbital Eccentricity}\label{sec:eexcit}

As stated above the $\PN$ effects may suppress the Newtonian Kozai-Lidov oscillations if the corresponding $\PN$ timescale is much smaller than the Newtonian  quadrupole one.
However, we identify two regimes where the combined secular Newtonian and $\PN$ effects produce interesting different behavior in three-body systems:
 (i) if the quadrupole order terms are comparable to or somewhat smaller than the $\PN$ ones, or (ii) for comparable-mass inner binaries where the $\PN$ effects are subdominant relative to the Newtonian quadrupole terms but they are comparable to the Newtonian octupole timescale.
We discuss these two regimes in detail below.

\subsection{Eccentricity Peak in the Restricted Three Body Problem for a Massive Perturber}

The standard lore says that GR effects (or $\PN$ effects in our case) suppress the eccentricity growth of the inner orbit in a three body system \cite[e.g.,][]{Bla+02,Dan}. However, \citet{Ford00}, studying the  triple system PSR~B1620$-$26, showed that the Newtonian octupole and the leading order  $\PN$ corrections [i.e.,~Eq.~(\ref{eq:1PNg1})]
 can lead to the excitation of the eccentricity of the inner orbit. We repeat and extend the investigations of that study for a broader range of three-body systems and examine (i) whether the new $\PN$ terms derived in \S\ref{Sec:Ham} give rise to different behavior; and (ii) whether Newtonian octupole terms can significantly change the evolution of the three-body system in the presence of the $\PN$ terms by producing flips of the inner orbit and eccentricities close to unity \citep{Naoz11,Naoz+11sec}.
We begin by exploring systems in which $m_3\gg m_1$ ($q_m\gg 1$), and investigate the opposite limit in the next subsection.

\begin{figure*}[t]
\begin{center}
\includegraphics[width=10cm,clip=true]{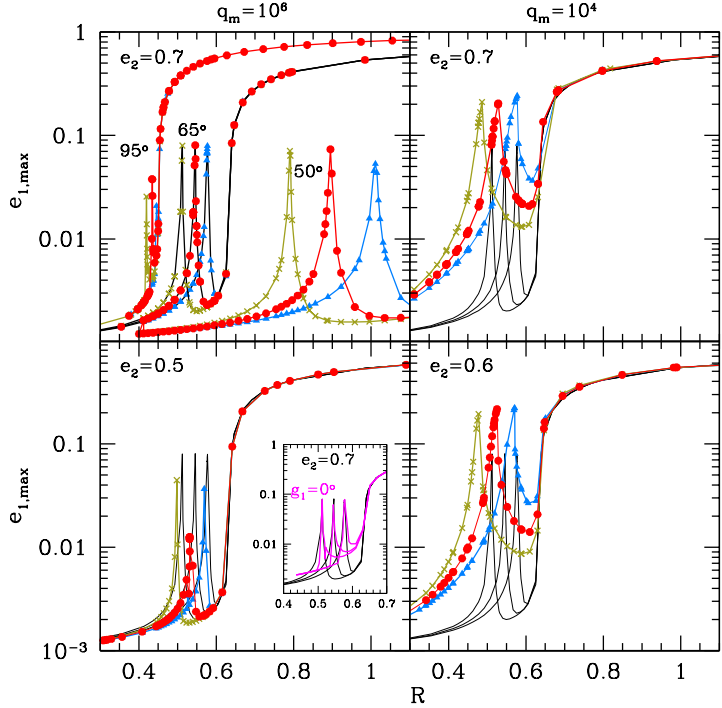}
\caption{The maximum eccentricity in a  triple system in the test-particle approximation for different mass ratios ($q_m=10^6$, left hand panels, and $q_m=10^4$, right hand panels) as a function of the relative timescales of the $\PN$ and secular Newtonian quadrupole effects ($\mathcal{R}$, see Eq.~\ref{eq:R}). We consider the $\PN$ evolution including terms only up to $\mathcal{O}(a_1^{-2})$, $\mathcal{O}(a_2^{-2})$ and the interaction term (blue triangles, green crosses, and red squares, respectively). We show three different initial outer orbital eccentricities: $e_2=0.7$ (top panels) $e_2=0.6$ and $0.5$, bottom right and left hand respectively. In all these examples, we set $m_1=1$~M$_\odot$, $m_2=0.001$~M$_\odot$, $m_3=q_m m_1$ and we vary both the inner and outer SMAs to match the different values of  $\mathcal{R}$, see Eq.~(\ref{eq:R}). The system is initialized with $e_1  =0.001$,  $g_2=0^\circ$ and $g_1=240^\circ$  (inset shows the results for initializing $g_1=0^\circ$ and $240^\circ$). In all panels we set the initial mutual inclination to $65^\circ$. In the top left panel, we also set the initial inclination to $95^\circ$ (left set of lines) and $50^\circ$ (right set of lines) . We compare the fiducial example ($q_m=10^6, i_\tot=65^\circ, e_2=0.7$), black lines in each panel, to systems with a different mass ratio ($q_m=10^4$), right hand panels), and different eccentricities, see labels top right panel. Observe the emergence of resonant-like eccentricity excitations.}  \label{fig:emax}
\end{center}
\end{figure*}

Let us systematically examine the parameters where the competition between the secular Newtonian Hamiltonian and $\PN$ corrections lead to the excitation of the inner orbital eccentricity as opposed to a suppression. We do this by preforming a large number of simulations starting from a nearly zero eccentricity for the inner binary and varying the following dimensionless parameter:
\begin{eqnarray}
\label{eq:R}
\mathcal{R}&=& \left.\frac{t^{\PN}_{a_1^{-2}}}{t^{\rm N}_{\rm quad}}\right|_{e_1=0}
=\frac{1}{3}\frac{(a_1 / R^g_1)^{4}}{(a_2/R^g_3)^3}\frac{1}{q_m^{2}(1-e_2^2)^{3/2}} \ .
\end{eqnarray}
where we have substituted Eqs.~(\ref{eq:tquad}) and~(\ref{eq:tGRa1}) with $e_1=0$. The quantity in Eq.~\eqref{eq:R} compares the timescales of the leading order $\PN$ and secular Newtonian effects. We find that a resonant--like eccentricity excitation can take place around $\mathcal{R}=1$ for zero initial eccentricity as shown below. Note, however, that another important ingredient to this resonant-like behavior is the Newtonian octupole term, which is most obvious in simulations with a low initial mutual inclination, for which the quadrupole approximation is subdominant (see also \S \ref{PSR} below and \citealt{Ford00}). In other words, neglecting the contribution of the octupole--level of approximation, one would miss entirely the resonant behavior.

Figure~\ref{fig:emax} shows the maximum eccentricity achieved during the course of the evolution of the system studied, after $1000$ quadrupole cycles\footnote{Each point in this figure corresponds to a separate, high-resolution three-body evolution with $\PN$ effects, each of which takes approximately $3$ days to complete per computer core. We also preformed convergence tests using longer integration times at high resolutions, and found that for this type of systems at least $1000$ quadrupole cycles are needed for convergence, over the parameter range considered in the Figure. } as a function of $\mathcal{R}$. We also examined the effects of various $\PN$ terms by repeating the calculations using $\bar{\Ham}_{\rm N}+\bar{\Ham}_{a_1^{-2}}^{\PN}$ (blue triangles), then adding $\bar{\Ham}_{a_2^{-2}}^{\PN}$ (green crosses), and finally including $\bar{\Ham}_{\rm int}^{\PN}$ (red circles). The fiducial example chosen (black lines in all panels) has $m_1=1$~M$_\odot$, $m_2=0.001$~M$_\odot$, and $m_3=10^6$~M$_\odot$ (i.e. $q_m=10^6$) with initial conditions $e_1=0.001$,  $e_2=0.7$, $i_\tot=65^\circ$, $g_1=240^\circ$ and $g_2=0^\circ$. To generate Fig.~\ref{fig:emax}, we used different runs and changed both $a_1$ and $a_2$. For a choice of mass ratio and initial mutual inclination the parameter $\mathcal{R}$ collapses the different runs with different SMAs into one curve.
Note that changing the initial value of the argument of periapsis does not change the location of the resonance, but it does change its width, as shown in the inset of Fig.~\ref{fig:emax}, where we compare the fiducial example with an initial $g_1=240^\circ$ to an initial $g_1=0$. The different panels show that the amplitude and width of the resonance are different for different $q_m$ and $e_2$, as they depend on the Newtonian octupole terms. The top left panel shows that different initial inclinations change both the location of the resonance and its width (as a function of $\mathcal{R}$), i.e., low initial inclination leads to broad resonances. The time evolution of the systems with parameters on the peak of the resonance of the $i_\tot=95^\circ$ and $65^\circ$ cases are shown in Fig.~ \ref{fig:e1_excitGR}.

To the left of the resonance peak, the maximum eccentricity converges asymptotically to the initial inner eccentricity for $\mathcal{R}\ll 1$ (i.e. in this case the eccentricity is not excited). This was chosen to be $e_1=10^{-3}$ in Fig.~\ref{fig:emax}. Decreasing the initial eccentricity (not shown) changes the figure at $\mathcal{R} \ll 1$, but it does not change the amplitude and location of the resonance and the behavior at larger $\mathcal{R}$. To the right of the resonance ($\mathcal{R}\gtrsim 1$), the maximum eccentricity starts to increase when the eccentric Kozai-Lidov mechanism begins to dominate. For $q_m\gg 1$ and $\mathcal{R}\gg 1$, we find asymptotically $e_1\to 1$ (see \S\ref{PSR} for small $q_m$).

A binary that slowly shrinks due to GW emission, slowly changes $\mathcal{R}$ and may sweep across the resonant regions of eccentricity excitation shown in Fig.~\ref{fig:emax}. Thus, resonant $\PN$ eccentricity increase may take place in many inspiraling systems some time during their evolution. Whether this occurs or not depends on the masses and SMAs; a detailed analysis is left to future work.

\subsection{The Case of PSR~B1620$-$26}\label{PSR}

Although the above discussion of the resonant behavior assumed a test particle ($m_2\sim 0$), general mass-ratio triples also exhibit a similar effect \citep[as was first considered by][]{Ford00}, provided $t^{\PN}_{a_1^{-2}}$ is shorter then $t^{\rm N}_{\rm quad}$ [see Eq.~(\ref{eq:tGRa1})].
To find $\mathcal{R}$ in the general case, we can simply set
\begin{equation}
q_m\equiv m_3/(m_1+m_2)
\end{equation}
in Eq.~(\ref{eq:R}). \citet{Ford00} observed a resonant-like eccentricity increase while studying the triple system PSR~B1620$-$26, which is located near the core of the globular cluster M4.  They showed that this resonant behavior may explain the unusually large eccentricity of the inner binary, which contains a millisecond radio pulsar of $m_1=1.4 \msun$ and a companion of $m_2=0.3\msun$ \citep{ML88}.

For completeness, we repeat and extend the calculation of \citet{Ford00} by fixing the inner orbit's SMA, changing the outer orbit's SMA, and including all $\PN$ terms.
We choose two different values for the inner binary to explore the sensitivity to these parameters. First, following  \citet{Ford00Pls}, we consider $a_1/ R_1=5.6\times 10^7$ (i.e. $a_1=0.77$~AU). Additionally, we consider $a_1/ R_1=3.6\times 10^8$ (i.e. $a_1=5$~AU). We adopt parameters for the outer perturber from \citet{Ford00Pls}: $m_3=0.01\, \msun$ and  $e_2=0.45$. We initialize the system with $g_1=g_2=0^\circ$, $e_1=10^{-4}$, and we also choose two different initial inclinations, $i_\tot=65^\circ$ and $i_\tot=0^\circ$. A mutual inclination of $i_\tot=0^\circ$ highlights that the perturbations of the outer orbit affect the inner orbit
due to the Newtonian octupole term, even far from the nominal Kozai-Lidov regime. In this configuration, the $\bar\Ham_{a_1^{-2}}^{\PN}$ term is the most significant $\PN$ effect, as shown in Figure \ref{fig:timescales} (see the rectangle in the top left panel).

\begin{figure}[tb]
\begin{center}
\includegraphics[width=8.5cm,clip=true]{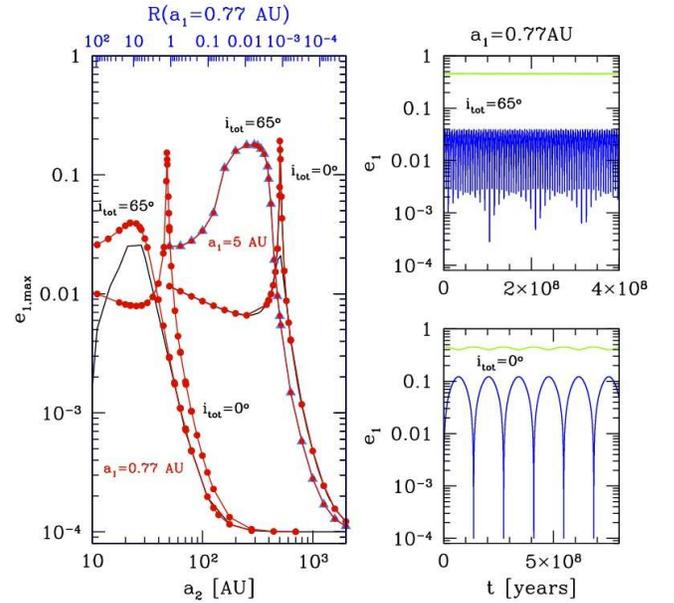}
\caption{The maximum eccentricity as a function of the outer orbit's SMA (left  panel) in a triple for the case of
PSR~B1620$-$26. The inner binary is a millisecond pulsar of mass
$1.4\,M_\odot$ with a companion of $m_2=0.3\msun$, and the outer
body has mass $m_3=0.01\msun$. The inner orbit has $a_1 = 5\,$AU and $0.77$~AU, in two different sets of simulations (see labels).
The initial eccentricities are $e_1 =10^{-4}$ and $e_2 =0.45$ and the initial
relative inclination $i_\tot=65^\circ$ and $i_\tot=0^\circ$.  The argument of pericenter of the inner and outer
orbits are initially set to zero. The top axis show the value of $\mathcal{R}$ for the case of $a_1=0.77$~AU, see equation (\ref{eq:R}). Note that the $i_\tot=0^\circ$ resonance happens when $\mathcal{R}\sim 1$. The thin black lines show the results of integrating the system over a single Kozai-Lidov cycle. The left thin black line is for the case of $a_1=0.77$~AU and $i_\tot=65^\circ$, while the right one is for $a_1 = 5\,$AU and  $i_\tot=0^\circ$. Although we include all $\PN$ terms described in previous sections, the curves corresponding to the lower order ones exactly overlap the curves including the interaction term
(i.e., the leading $\PN$ term is the most dominant in the evolution of the system).
In the right panels, we show the time evolution of the eccentricity of the inner (blue line) and outer binary (green line). Here we set $a_1=0.77$~AU $i_\tot=0^\circ$,  $a_2=48$~AU in the bottom and $i_\tot=65^\circ$, $a_2=22$~AU in the top panel, respectively. In this case, the resonant eccentricity excitation due to the $\PN$ terms  reaches higher values then the one achieved by the eccentric Kozai-Lidov mechanism for small (large) $a_2$ ($\mathcal{R}$).
 }\label{fig:e1PSR}
\end{center}
\end{figure}

The left-panel of Fig.~\ref{fig:e1PSR} shows the inner orbit's maximum eccentricity as a function of the outer orbit's SMA (or equivalently the  $\mathcal{R}$ value for $a_1=0.77$~AU). This figure confirms the resonant--like increase in eccentricity found in \citet{Ford00} (their Figure 14). This figure also shows that the resonant-like eccentricity increase is present for a large range of $a_{2}$ values, even for systems where the excitation of the eccentricity due to the Newtonian octupole term is somewhat suppressed, due to comparable masses for the inner orbit \citep[e.g.,][]{Naoz+11sec}. Although changes in the outer orbit's eccentricity (see the right panels of Fig.~\ref{fig:e1PSR}) do not change the location of the resonant peaks, their amplitude does change for large inclinations. The $a_1=5$~AU case has irregular behavior and results in a higher inner and outer orbital eccentricity. Note that $\Ham_{a_1^{-2}}^{\PN}$ term is the dominant one here and the other $\PN$ terms
are negligible in this configuration, as can be see from the black rectangle in the top left panel in Figure \ref{fig:timescales}.

The right panels of Fig.~\ref{fig:e1PSR} show the time evolution of the inner and outer orbital eccentricity for the $a_1=0.77$~AU case with $i_\tot=0^\circ$ and $a_2/R_3=4.9\times 10^{11}$ ($a_2=48$~AU, bottom panel) and $i_\tot=65^\circ$ and $a_2/R_3=2.2\times 10^{11}$ ($a_2=22$~AU, top panel). Unlike the systems considered in Fig.~\ref{fig:emax}, the outer orbit's eccentricity oscillates slightly (see right hand panels in Figure \ref{fig:e1PSR}). Furthermore, the eccentricity peak is even larger than the eccentricity reached for $\mathcal{R}\gtrsim 1$, which shows that the $\PN$ terms further increase the inner orbit's eccentricity above the excitation induced by the Newtonian eccentric Kozai-Lidov mechanism. This was not the case in Figure~\ref{fig:emax} primarily because $q_m\gg 1$ there.

Although we integrated the system for up to $1000$ Kozai-Lidov cycles, the eccentric Kozai-Lidov process did not seem to induce chaotic behavior in this configuration (since $m_3<m_1+m_2$).  Numerical convergence was reached already after a few hundred Kozai-Lidov cycles of evolution (in thin black lines we show the results of integrating the system up to 1 Kozai-Lidov cycle). Note that the this system was integrated both using the calculation presented here and using direct 3-body in \citet[][figure 15]{Ford00}; here we find perfect agreement with their results.  

The top right panel of Fig.~\ref{fig:e1PSR} shows that integrating over only one Kozai-Lidov cycle, for the $i_\tot=65^\circ$ case, misses the long timescale oscillation, since the largest eccentricity is reached only after several Kozai time scales. This explains why our eccentricity peaks are slightly higher than those of \citet{Ford00}, as we see for the  $i_\tot=65^\circ$ case in the left panel. This also explains the somewhat larger $e_{1,\rm max}$ values we found compared to \citet{Ford00} in the regime where the $\PN$ effects are subdominant and the eccentricity of the inner orbit increases due to the eccentric Kozai-Lidov mechanism. Note however, that the lifetime of this system in the core of the globular M4 is about one Kozai-Lidov cycle, which explains why \citet{Ford00} did not bother to evolve over many Kozai-Lidov cycles.

\subsection{Orbital Flips and Eccentricity Excitation for Comparable-mass Inner Binary}

\begin{figure}[tb]
\begin{center}
\includegraphics[width=8.5cm,clip=true]{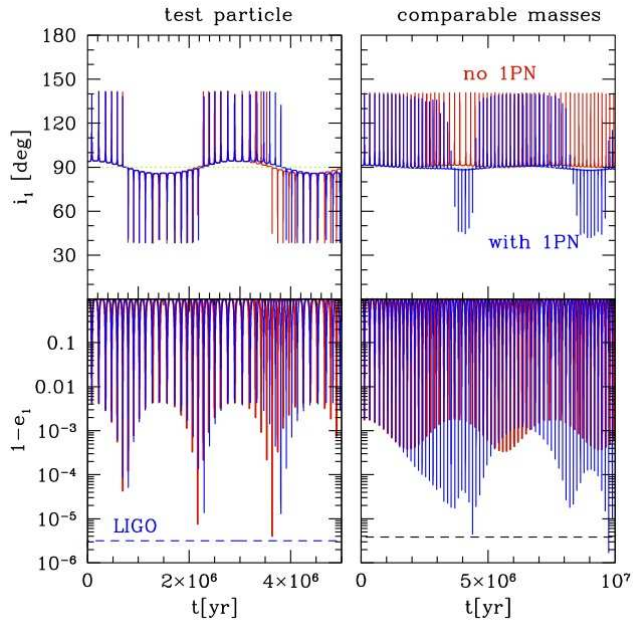}
\caption{
Eccentricity excitation and orbital flips for the Newtonian octupole and the 1PN approximations (up to the interaction term, see text for details) as a function of time. We compare the test-particle case ($m_2\to 0$ left panels) to a comparable mass case ($m_2=8$~M$_\odot$ right panels) in the inner binary with $m_1=10\,M_\odot$, always with an outer object of mass $m_3 = 30\,M_\odot$. The separation of the inner orbit is $a_1/R^g_1=1\times 10^8$ (corresponding to $10$~AU and orbital period $P_1\sim 10\,$yr), and the outer orbit's separation is $a_2/R^g_3=1.69\times 10^9$ (corresponding to
502~AU and $P_2=2.8 \times 10^3\,$yr). The initial eccentricities are $e_1 =0.001$ and $e_2 =0.7$ and the initial relative inclination $i_\tot=94^\circ$.  The arguments of pericenter of the inner and outer orbits are initially set to $240^\circ$ and zero respectively. For these examples $\mathcal{R}\gg 1$. We show with red lines evolutions {\it without} $\PN$ corrections (curves including only the lower order $\PN$ terms simply overlap this).
We also show the minimum eccentricity corresponding to the detectable LIGO frequency range (horizontal lines in the bottom panels).
The 1PN corrections help to further increase the eccentricity and lead to orbital flips for the inner binary for comparable masses.
} \label{fig:e1_excite}
\end{center}
\end{figure}

If the inner binary consists of comparable mass-objects, the Newtonian octupole term is suppressed (see Eq.~\ref{eq:epsiM}).  Recently, \citet{Sharpee+12} considered the evolution of triple systems with comparable masses and showed that the eccentric Kozai-Lidov evolution can be triggered if one of the stars in the inner binary loses mass. We show here that the eccentric Kozai-Lidov evolution can also be triggered without mass loss, but accounting for $\PN$ effects, as shown in Fig.~\ref{fig:e1_excite}. For this figure, we set $m_1=10\,M_\odot, m_3 =30\,M_\odot$, $a_1/R^g_1=1\times 10^8$ (corresponding to $10$~AU), and $a_2/R^g_3=1.7\times 10^9$ (corresponding to $502$~AU). The initial eccentricities were $e_1=0.001$ and $e_2 =0.7$ and the initial relative inclination $i_\tot=94^\circ$.  The argument of pericenter of the inner and outer orbits was initially set to $240^\circ$ and zero, respectively. The two panels in this figure differ in the choice of $m_{2}$, i.e.~in the left panel $m_2 = 0.001$ and in the right panel $m_{2} = 8 M_{\odot}$. While a test-particle evolution is relatively insensitive to the $\PN$ terms in this case,
comparable mass systems present qualitatively different behavior. In particular, while the Newtonian eccentric Kozai-Lidov effect is suppressed for comparable masses, $\PN$ effects can trigger it.
A possible reason for the qualitative difference is that changing $m_2$ from zero to $8$~M$_\odot$ resulted in a configuration for which the leading order $\PN$ timescale is closer to (but still slightly longer than) the octupole time scale. In the test particle case, the octupole timescale is two orders of magnitude shorter than the shortest $\PN$ timescale.

\begin{figure*}[htb]
\begin{center}
\includegraphics[width=9.5cm,clip=true]{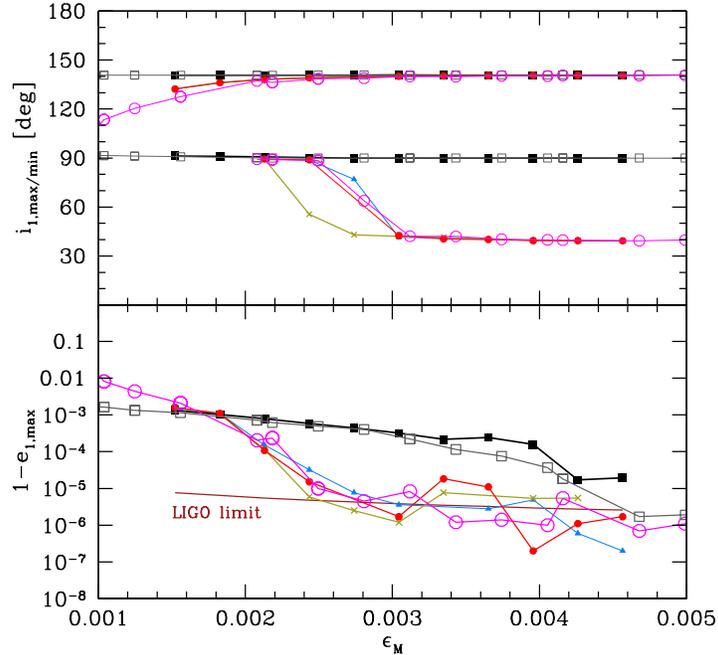}
\caption{\label{fig:e1_excite2}Excitation of the eccentricity and inclination due to $\PN$ effects for Kozai-Lidov timescales much shorter than the lowest $\PN$ timescales (as parametrized through $\epsilon_M$, see Eq.~\ref{eq:epsiM}). We examine the maximum eccentricity reached, and plot $1-e_{1,{\rm max}}$ (bottom panel), and the maximum and minimum inclination reached in the runs (top panel). We show two examples.
The first system is initialized with $m_1=10$~M$_\odot$, $m_2=8$~M$_\odot$ $m_3=30$~M$_\odot$  and $e_2=0.7$. We initialize the system with  $e_1=0.001$, $g_1=240^\circ$, $g_2=0$ and $i_\tot=94^\circ$. We vary both the inner and outer SMAs to match the different values of  $\epsilon_M$ depicted in the figure. We consider the Newtonian approximation (filled black squares) and the three $\PN$ level of approximations, $\mathcal{O}(a_1^{-2})$, $\mathcal{O}(a_2^{-2})$ and the interaction term (filled blue triangles, green cross and filled red squares, respectively).  In the second system, we set $m_1=1$~M$_\odot$, $m_2=1.2$~M$_\odot$ $m_3=3$~M$_\odot$  and $e_2=0.7$.
We initialize the system with  $e_1=0.001$, $g_1=0^\circ$, $g_2=0$ and $i_\tot=94^\circ$. We vary both the inner and outer SMAs to match the different values of  $\epsilon_M$ depicted in the figure. For this example, we consider the Newtonian approximation (empty gray squares) and up to the interaction level of the $\PN$ approximation (empty magenta circles).
We also show the detectable LIGO frequency limit for the first example, where we set $a_2=501$~AU and varied $a_1$ between $5$~AU to $15$~AU (solid brown line).
}
\end{center}
\end{figure*}

The excitation of the eccentricity depends on the importance of the Newtonian octupole term. Without $\PN$ effects, comparable mass triples result in $\epsilon_M\to 0$ [Eq.~(\ref{eq:epsiM})], which suppresses the eccentricity excitations and the flipping of the inner orbits. This can be seen in Fig.~\ref{fig:e1_excite2} (black and gray lines) for two examples, where we consider $\Delta m=|m_1-m_2|=2$~M$_\odot$,  (black lines) and  $\Delta m=0.2$~M$_\odot$ (gray lines). For both of these, $m_3=3 m_1$ and the two systems were initialized with $e_2=0.7$, $e_1=0.001$ and $i_\tot=94^\circ$. However, we find that although $\PN$ effects are small during a single orbit, they can be significant on much longer timescales, and can lead to significant eccentricity growth and orbital-flips\footnote{The choice of argument of periapsis does not change the outcome; the first example was initialized with $g_1=240^\circ$ while the second had $g_1=0^\circ$, both had $g_2=0^\circ$.}.  The top panel shows that for $\epsilon_M \geq 0.002$  a flip is triggered due to the $\PN$ terms. In other words,  for $\epsilon_M \geq 0.002$  the colored  curves deviate (including $\PN$ effects) from the black and grey ones (Newtonian effects only).
For larger $\epsilon_M$, where the eccentric Kozai-Lidov mechanism is triggered, the maximum eccentricity can be very close to unity, and thus due to the chaotic nature of the system, the maximum eccentricity shown should be considered as a lower limit.  For  $\epsilon_M\gg0.005$ we could not reach convergence after $1000$ quadrupole cycles, since the eccentricity is very close to unity (see Teyssandier et al; in preparation).

In the above examples, a requirement for eccentricity excitation is that the $\PN$ timescale, $t^{\PN}_{a_1^{-2}}$, be shorter than or comparable to the octupole timescale $t^{\rm N}_{\rm oct}$, [see Eq.~(\ref{eq:toct})], i.e., $t^{\rm N}_{\rm quad}\lesssim t^{\PN}_{a_1^{-2}} \sim t^{\rm N}_{\rm oct}$. A possible explanation for the excitation of the eccentricity in these cases is the following. Neglecting $\PN$ effects, comparable-masses in the inner binary suppress the Newtonian octupole effects and the outer potential is effectively quadrupolar. However, GR precession of the inner orbit breaks this symmetry. As long as GR precession occurs on a timescale comparable to (or slightly smaller than) the octupole one, the eccentric Kozai-Lidov mechanism is then triggered.

Let us now discuss the implications of these finding for direct GW detections using Earth-based instruments like LIGO and VIRGO. The characteristic frequency of the GW signal is $f_p=v_p/r_p$, where $v_p$ and $r_p$ are the orbital velocity and radius at pericenter \citep{Peters64}. Thus, $f_p = 2 \pi (1+e)^{1/2} (1-e)^{-3/2} P^{-1}$, where $P$ is the orbital period. We assume that the GW signal is in the detectable frequency band if $f_p > 5~{\rm Hz}$. For a comparable-mass inner binary, the Newtonian eccentric Kozai-Lidov mechanism is suppressed\footnote{Although suppressed, note that the eccentricity can still reach 0.999 in this case.} and the eccentricity remains smaller than in the test particle case.
The GW frequency emitted by a circular binary with an orbital period larger than a second is too small for a LIGO detection. However, the $\PN$ eccentricity excitations discussed in this paper lead to a much larger $f_p$ and might lead to GWs in the LIGO band.
In particular, Figs.~\ref{fig:e1_excite} and \ref{fig:e1_excite2} show examples where $1-e_1$ can be as small as $10^{-5}$ to $10^{-6}$ for a comparable-mass inner binary. These sources enter the LIGO GW frequency band if their orbital period is less than 1 to 60 years.
However, note that this estimate is oversimplified because it neglects the backreaction of GW emission on the evolution. The latter strongly reduces the SMA of the binaries during close approaches, and may lead to an eccentric inspiral and merger within a Kozai-Lidov period. The GW inspiral may deliver the binaries to the LIGO frequency band even if the signal is outside the LIGO band during the Kozai-Lidov oscillations.
If the event rate of these sources is sufficiently large within the LIGO detection range, these sources could constitute a distinct population for LIGO \citep{Wen,AP12}.

\section{Discussion}\label{sec:dis}

The Kozai-Lidov mechanism  \citep[][see below]{Kozai,Lidov}, has been shown to play an important role for highly inclined hierarchical triples, from planetary systems to stellar size and/or massive compact  objects \citep[e.g.,][and references therein]{Naoz+11sec}. For an eccentric outer perturber, the eccentricity of the inner orbit can reach values extremely close to unity, and the inclination can flip from prograde to retrograde \citep{Naoz11,Naoz+11sec}. The quadrupole Kozai-Lidov oscillations between the eccentricity and inclination still persist at octupole order, but they are further modulated on long timescales.

We have here studied how the Kozai-Lidov mechanism is affected by $\PN$ corrections to the three-body Hamiltonian, focusing on secular and hierarchical three body systems.
We expanded the $\PN$ Hamiltonian in the ratio of SMAs ($\alpha$) to third order beyond leading, i.e.~the leading-order terms in the $\PN$ Hamiltonian perturbation scale here as $a_{1}^{-2}$ and we carried out an expansion up to relative ${\cal{O}}(\alpha^{3})$.
We also averaged over the orbital timescale of the inner and outer binary to investigate the long-term secular evolution of the system (\S \ref{Sec:Ham}). We examined the effects of the different $\PN$ terms in this expansion: $\PN$ precession of the inner orbit due to $\bar{\Ham}_{a_1^{-2}}^{\PN}$ (Eq.~\ref{eq:1PNa1}); $\PN$ precession of the outer orbit due to $\bar{\Ham}_{a_2^{-2}}^{\PN}$ (Eq.~\ref{eq:1PNa2}); and a new $\PN$ interaction term between the two orbits,  ${\bar\Ham}_{\rm int}^{\PN}$ (Eq.~\ref{eq:1PNint}), which introduces a new inclination and eccentricity dependent modulation (e.g., Fig.~\ref{fig:e1_excitGR}).

We compared the different timescales associated with the secular Newtonian and different $\PN$ terms  (see Fig.~\ref{fig:timescales}). If the timescales associated with the $\PN$ effects are much shorter than the timescales associated with the eccentric Kozai-Lidov mechanism, i.e.~the secular Newtonian timescales, the growth of the eccentricity in the inner orbit tends to be suppressed. We confirm that the excitation of the eccentricity is indeed suppressed for systems where the Kozai-Lidov timescale is many orders of magnitude longer than the $\PN$ timescales. However, if the timescales of the $\PN$ effects are comparable to the secular Newtonian ones (see Fig.~\ref{fig:timescales}), we found two interesting regimes that present qualitatively different behavior.

The first regime is where the $\PN$ timescales are comparable but slightly shorter than the Newtonian Kozai-Lidov timescale. \citet{Ford00}, studying the PSR~B1620$-$26 triple system, noted that the inner eccentricity may be greatly increased around some critical value of the outer SMA, due to the $\bar{\Ham}_{a_1^{-2}}^{\PN}$ term and the octupole term. We extended this calculation by including all averaged $\PN$ terms up to $\mathcal{O}(\alpha^3)$ and the Newtonian octupole term \citep{Naoz+11sec}, as well as exploring a wide region of phase space. We confirmed \citet{Ford00} result and found a resonant-like behavior, where the inner orbital eccentricity is greatly increased compared to the Newtonian case. This behavior exists also when including  all averaged $\PN$ terms and for a wide range of mass ratios and orbital parameters. We parameterized the location of the resonant peak in terms of the SMAs by defining a parameter, $\mathcal{R}$ in Eq.~(\ref{eq:R}), as the ratio of the leading-order $\PN$ and secular Newtonian terms. This parameter depends on the ratio of the mass of the outer perturber to the total mass of the inner binary. The presence of the octupole term is important for the resonant $\PN$ eccentricity excitation, which is most apparent in the examples with a small mutual inclination. For systems where either the inner or the outer binary shrinks, for example due to GW radiation-reaction, the triple may pass through this three-body $\PN$ resonance. The amplitude and location of the resonance changes due to $\PN$ terms as a function of $\mathcal{R}$. We found that lower mutual inclinations in the prograde regime cause a wider peak (in terms of $\mathcal{R}$), while a less massive outer body tends to produce wider and higher amplitude peaks. A detailed investigation of the properties of the resonance is beyond the scope of this paper, but could be the subject of future investigations.

It is important to note that the outer orbit precession and the interaction term affect the overall time evolution  (see  Figure \ref{fig:e1_excitGR}). Since these terms are a result of the expansion of the three body $\PN$ Hamiltonian in  $\alpha$, it is not surprising that the different terms affect the location of the resonant like behavior
 (e.g., Figure \ref{fig:emax}). It is interesting however, that they produce a qualitatively different time evolution of the system (e.g., bottom panels of Figure \ref{fig:e1_excitGR}).   This suggests that a system evolved under GR effects in the {\bf presence} of a third body has richness to it that should be examined in more detail. This is the subject of future investigation in the framework of direct 3-body integration. 

The second regime that exhibits qualitatively different behavior from that obtained with a quadrupole Newtonian Kozai-Lidov treatment is when the quadrupolar secular Newtonian timescales are shorter than the $\PN$ ones and when the inner binary has comparable mass components. The eccentric Kozai-Lidov mechanism, neglecting $\PN$ effects, is suppressed when $m_1\to m_2$, since the outer orbit's potential is effectively quadrupolar. As we showed in this paper, $\PN$ effects can break symmetry and excite eccentricity, triggering the eccentric Kozai-Lidov mechanism. As long as $\PN$ precession occurs on a comparable timescale (or lower) than the Newtonian octupole precession, i.e, $t^{\rm N}_{\rm quad}\lesssim t^{\PN}_{a_1^{-2}}\sim t^{\rm N}_{\rm oct}$, the eccentric Kozai-Lidov mechanism will be triggered.

Eccentricity excitations are particularly interesting in the context of possible GW detections \citep{Wen,2010PhRvD..81b4007B,2005ApJ...634..921A,2010ApJ...719..851S}. If such excitations were not present, the frequency of the GWs emitted by the inner binary would be typically too low for detection with LIGO \citep[see however][for eccentric binaries which form in the LIGO band]{2009MNRAS.395.2127O,2011arXiv1109.4170K}. However, if eccentricity is secularly excited through a three-body interaction, the frequency of the GWs is also increased during pericenter passage, thus bringing the signals into the detector's sensitivity band. Such large eccentricities would then lead to GW-driven inspiral and the eventual merger of binaries. Whether such eccentric signals can be detected or not will depend on how close such sources are to Earth. But if detections are made with sufficiently high signal-to-noise ratio, then GWs could be used to measure the eccentricity of the inner binary, and thus, distinguish between different source populations.

\section*{Acknowledgments}
We thank Alessandra Buonanno, Fred Rasio and Gongjie Li for useful discussions, and we also thank Cole Miller for carefully reading the first draft of the paper and sending useful comments. We thank our anonymous referee for useful remarks. We  thank Yoram Lithwick for the use of his allocation time on the computer cluster Quest. This research was supported in part through the computational resources and staff contributions provided by Information Technology at Northwestern University as part of its shared cluster program, Quest. SN acknowledge partial supported by NASA through a Einstein Post- doctoral Fellowship awarded by the Chandra X-ray Center, which is operated by the Smithsonian Astrophysical Observatory for NASA under contract PF2-130096. This work was supported in part by NSF grant PHY-1114374 and AST-0907890, as well as NASA grants NNX08AL43G and NNA09DB30A and NNX11AI49G. BK acknowledges support from NASA through Einstein Postdoctoral Fellowship Award Number PF9-00063 issued by the Chandra X-ray Observatory Center, which is operated by the Smithsonian Astrophysical Observatory for and on behalf of the National Aeronautics Space Administration under contract NAS8-03060. NY also thanks the Institute for Theory and Computation at the Harvard Smithsonian Center for Astrophysics for their hospitality.

\appendix
\section{A. Two body Systems - Effective One Body}\label{sec:2body}

GR pericenter precession has been studied in great detail and used to test Einstein's theory in the Solar System, for example with observations of the perihelion precession of Mercury \citep[e.g.,][]{Shapiro+72}. The simplest method to derive such precession is to consider test-particle motion in an effective potential, assuming that GR introduces small corrections to Newtonian dynamics and small eccentricities  \citep[e.g.,][chapter 25 p.~668--670]{Gravitation}\footnote{Note that the same precession rate can be also derived directly from the $\PN$ metric  \citep[e.g.,][chapter 40 p.~1100-1112]{Gravitation}.}. The Hamiltonian
\citep[e.g.][]{Artemova+96,MH02} is simply derived by integrating over the precession rate. Although this Hamiltonian leads to the correct ISCO location, if one uses the full expansion given in \citet[][Eq.~ 4]{Artemova+96}, it is not equal to the $\PN$ Hamiltonian.

The purely orbital (non-spinning) 3PN Hamiltonian was derived in \citet{JS98,JS01E} (in the center of mass frame, and after subtraction of the total rest-mass term). Here, we focus only on expansions to $\PN$ order for a two body system (with masses $m_1$ and $m_2$ and momenta $p_1$ and $p_2$, respectively). The Hamiltonian is then \citep[e.g.,][]{Buonanno+06}:
\begin{eqnarray}
\label{eq:1PN2b}
\Ham^{2\rm body}_{1PN}&= &-  \frac{(m_1^3+m_2^3) p^4}{8 c^2 m_1^3 m_2^3}   - \frac{k^2 (3 m_1^2+ 7m_1 m_2 +3 m_2^2) p^2}{ 2c^2m_1 m_2 r} - \frac{k^2 ({\bf p}\cdot {\bf r})^2 }{2 c^2 r^3} + \frac{k^4 (m_1+m_2)^2 \mu }{2 c^2 r^2} \ , \nonumber
\end{eqnarray}
where ${\bf r}$ is the radius vector between the two bodies, with magnitude $r$, the linear momentum of the effective one body problem is simply $p=-p_1=p_2$ and $\mu=m_1m_2/(m_1+m_2)$. Eliminating the short-period terms in the Hamiltonian, using the Von Zeipel transformation \citep[for more details, see][]{bro59} for an orbit with SMA $a$ and eccentricity $e$, the double average Hamiltonian is given by
\begin{equation}
\label{eq:1PN2body}
\bar{\Ham}^{2\rm body}_{\PN} = \frac{k^4 \mu_{\inner} \left(15 {m_1}^2+29 {m_1} {m_2}+15 {m_2}^2\right)}{8 {a}^2 c^2 }-\frac{3k^4 {m_1} {m_2} ({m_1}+{m_2})}{{a}^2 c^2 \sqrt{1-{e}^2}} \ ,
\end{equation}
which is the same as Eq.~(\ref{eq:1PNa1}).

\section{B. The Von Zeipel transformation}\label{sec:VZ}

The technique, known as the Von Zeipel transformation
\citep[for more details, see][]{bro59} is being used in order to eliminate the
short-period terms in the Hamiltonian that depend of $l_1$ and $l_2$.
The technique had been used to derive the double average hierarchical three body Hamiltonian \citep[e.g.][]{Kozai,Har68,Har69,KM99,Naoz+11sec}. Here the Hamiltonian we consider  is simply $\Ham_{\tot,\PN}=\Ham_{\rm N}+\Ham_{\PN}$ (see \S \ref{Sec:Ham}). Following \citet{Naoz+11sec} Appendix A, we replace $\Ham$ by our $\Ham_{\tot,\PN}$. The equivalent of Equation (A7) at  \citet{Naoz+11sec}  is simply:
\begin{equation}
  \Ham_{\tot,\PN} = \Ham_1^K + \Ham_2^K + \Ham^{\rm N}_2+\Ham_{\PN},
\end{equation}
where $\Ham_1^K$ and $\Ham_2^K$ are the Kepler Hamiltonians that
describe the inner and outer  {\it{Newtonian}} orbits in the triple system, $\Ham^{\rm N}_2$ describes the {\it{Newtonian}} quadrupole interaction between the orbits (for the octupole interaction one can simply a $\Ham^{\rm N}_3$), and  $\Ham_{\PN}$ describes the  $\PN$ correction up to ${\cal{O}}(\alpha^{3})$. 
In this technique, we use a canonical transformation that can eliminate the $l_1$ and $l_2$ terms
from the $\Ham^{\rm N}_2+\Ham_{\PN}$ parts (which depends on $l_1$ and $l_2$), where the momenta are $p_i \in \left\{L_i, G_i, H_i \right\}$, and the
coordinates are $q_i \in \left\{l_i, g_i, h_i\right\}$.
Replacing $\Ham_2$ from \citet{Naoz+11sec} Appendix A, with $\Ham_2^{\rm tot}= \Ham^{\rm N}_2+\Ham_{\PN}$, we find (after following their derivation) the equivalent of their equation (A22)
\begin{equation}
  \Ham^{{\rm tot},*}_2\left( q_i^*, p_i^* \right)  =
  \frac{1}{4\pi^2} \int_{0}^{2\pi} d l_1^* dl_2^*\, \Ham^{\rm tot}_2\left( q_i^*,
    p_i^* \right) \  ,
\end{equation}
where the new momenta and coordinates have a superscript asterix. Since Hamiltonian is an additive quantity, and integral is an additive operation the overall new Hamiltonian after the canonical transformation is simply: 
\begin{equation}
  \Ham^{{\rm tot},*}_2  =
  \frac{1}{4\pi^2}\left(  \int_{0}^{2\pi} d l_1^* dl_2^*\, \Ham^{\rm N}_2 +  \int_{0}^{2\pi} d l_1^* dl_2^*\, \Ham^{\PN}  \right)\  .
\end{equation}
Therefore we can simply use the double averaged Newtonian Hamiltonian  derived in \citet{Naoz+11sec} and separately derive the double averaged  $\PN$ Hamiltonian.

\section{C. Equation of motions for the 1PN interaction}\label{sec:1PNint}

Using the canonical relations [eqs.~(\ref{eq:Canoni})], we find the equations of motion for the interaction part of the $\PN$ Hamiltonian:
\begin{eqnarray}
\label{eq:1PN3bdg1}
\frac{dg_1}{dt} \bigg |_{\PN({\rm int})} &=& \frac{k^4  m_1m_2 m_3 }{16 a_2^3 c^2 (1-e_2^2)^{3/2} (m_1+m_2)^2 (m_1+m_2+m_3)}\bigg\{ \frac{a_1}{G_1} (m_1+m_2+m_3) [ (1 - e_1^2) (  5m_1^2 -3 m_1m_2 + 5 m_2^2) \nonumber \\
&-&  9 \mathit{f}_{m_1m_2}  ( (1 - e_1^2) \cos 2g_1 +  2 \cos^2 i_\tot \sin^2 g_1)] +\frac{1}{G_2} [-8 \mathit{f}_{LL}  + \mathit{f}_i ] \bigg\}
\end{eqnarray}
where $\tilde{L}_{1,2}=L_{1,2}/\mu_{in,out}$, 
\begin{equation}
\mathit{f}_{m_1m_2} =  m_1^2 +  m_1m_2 + m_2^2 \ ,
\end{equation}
\begin{equation}
\mathit{f}_{LL} = \tilde{L}_1 \tilde{L}_2(m_1+m_2)  (4 (m_1+m_2)+3 m_3)  \ ,
\end{equation}
\begin{equation}
\mathit{f}_{e_1} = (2 - 5 e_1^2) m_1^2 +  3 (-2 + e_1^2) m_1 m_2 + (2 - 5 e_1^2)m_2^2 \ ,
\end{equation}
and also
\begin{equation}
\mathit{f}_i=3 a_1 (m_1+m_2+m_3) \cos i_\tot ( \mathit{f}_{e_1}+ 3 e_1^2 \mathit{f}_{m_1m_2}  \cos 2g_1 )
\end{equation}

\begin{eqnarray}
\frac{dg_2}{dt}&=& -\frac{k^4  m_1m_2 m_3 }{16 a_2^3 c^2 (1-e_2^2)^{3/2} (m_1+m_2)^2 (m_1+m_2+m_3)}\bigg\{ \frac{1}{G_1} [ 8 \mathit{f}_{LL}  -\mathit{f}_i] \\
&-& \frac{1}{2 G_2} [2 \cos i_\tot (-8  \mathit{f}_{LL} +\mathit{f}_i ) - 16 (m_1+m_2)^2 \cos i_\tot  \tilde{L}_1 \tilde{L}_2  [16 (m_1+m_2)  \tilde{L}_1 \tilde{L}_2 (7 (m_1+m_2)+6 m_3)\cos i_\tot  \nonumber\\
&+&\frac{3}{2}a_1(m_1+m_2+m_3) (- \mathit{f}_{e_1} [1+3\cos 2i_\tot]+18 e_1^2 \mathit{f}_{m_1m_2}  \cos 2 g_1 \sin^2 i_\tot )]  \bigg \}\nonumber
\end{eqnarray}

\begin{equation}
\frac{de_1}{dt} \bigg |_{\PN({\rm int})} =\frac{9 a_1 e_1 \sqrt{1-e_1^2} k^4  m_1 m_2 (m_1^2 + m_1 m_2 + m_2^2) m_3 \sin^2 i_\tot \sin (2 g_1)}{16 a_2^3 c^2 (1-e_2^2)^{3/2} L_1 (m_1+m_2)^2} \ .
\end{equation}
The change of the inner orbital angular momentum is simply
\begin{equation}
\frac{dG_1}{dt} \bigg |_{\PN({\rm int})} =\frac{9 a_1 e_1^2 k^4  m_1 m_2 (m_1^2 + m_1 m_2 + m_2^2) m_3 \sin^2 i_\tot \sin (2 g_1)}{16 a_2^3 c^2 (1-e_2^2)^{3/2}  (m_1+m_2)^2} \ ,
\end{equation}
while for the outer orbit it is simply zero. Thus,
\begin{equation}
  \frac{d H_1}{dt} \bigg |_{\PN({\rm int})} = \frac{\sin i_2}{\sin i_{\tot}} \frac{d G_1}{dt}  \ ,
\end{equation}
i.e.,
\begin{equation}
  \frac{d H_1}{dt} \bigg |_{\PN({\rm int})} = \frac{\sin i_2}{\sin i_{\tot}} \frac{9 a_1 e_1^2 k^4  m_1 m_2 (m_1^2 + m_1 m_2 + m_2^2) m_3 \sin^2 i_\tot \sin (2 g_1)}{16 a_2^3 c^2 (1-e_2^2)^{3/2}  (m_1+m_2)^2} \ ,
\end{equation}
The inclinations evolve according to $\dot{(\cos i_1)}= {\dot{H}_1} / {G_1} - \dot{G}_1 / {G_1} \cos i_1$ \citep[e.g.][]{Naoz+11sec}, and thus,
\begin{eqnarray}
\dot{(\cos i_1)} \bigg |_{\PN({\rm int})}& =& \frac{9 a_1 e_1^2 k^4  m_1 m_2 (m_1^2 + m_1 m_2 + m_2^2) m_3 \sin^2 i_\tot \sin (2 g_1)}{16 a_2^3 c^2 (1-e_2^2)^{3/2}  (m_1+m_2)^2} \\
&\times&  \frac{1}{G_1} \left(\frac{\sin i_2}{\sin i_{\tot}}  -\cos i_1 \right) \ , \nonumber
\end{eqnarray}
and since $\dot{(\cos i_2)}= {\dot{H}_2} / {G_2} - \dot{G}_2 / {G_2} \cos i_2$ and $\dot{H}_2=- \dot{H}_1$ \citep[e.g.][]{Naoz+11sec} we find
\begin{equation}
\dot{(\cos i_2)}=   \frac{\sin i_2}{G_2 \sin i_{\tot}} \frac{9 a_1 e_1^2 k^4  m_1 m_2 (m_1^2 + m_1 m_2 + m_2^2) m_3 \sin^2 i_\tot \sin (2 g_1)}{16 a_2^3 c^2 (1-e_2^2)^{3/2}  (m_1+m_2)^2} \ ,
\end{equation}

\bibliographystyle{hapj}

\bibliography{KozaiGR}

\begin{thebibliography}{78}
\expandafter\ifx\csname natexlab\endcsname\relax\def\natexlab#1{#1}\fi

\bibitem[{{Amaro-Seoane} {et~al.}(2012{\natexlab{a}}){Amaro-Seoane}, {Aoudia},
  {Babak}, {Binetruy}, {Berti}, {Bohe}, {Caprini}, {Colpi}, {Cornish},
  {Danzmann}, {Dufaux}, {Gair}, {Jennrich}, {Jetzer}, {Klein}, {Lang}, {Lobo},
  {Littenberg}, {McWilliams}, {Nelemans}, {Petiteau}, {Porter}, {Schutz},
  {Sesana}, {Stebbins}, {Sumner}, {Vallisneri}, {Vitale}, {Volonteri}, \&
  {Ward}}]{eLISA1}
{Amaro-Seoane}, P. {et~al.} 2012{\natexlab{a}}, ArXiv e-prints, 1202.0839

\bibitem[{{Amaro-Seoane} {et~al.}(2012{\natexlab{b}}){Amaro-Seoane}, {Aoudia},
  {Babak}, {Bin{\'e}truy}, {Berti}, {Boh{\'e}}, {Caprini}, {Colpi}, {Cornish},
  {Danzmann}, {Dufaux}, {Gair}, {Jennrich}, {Jetzer}, {Klein}, {Lang}, {Lobo},
  {Littenberg}, {McWilliams}, {Nelemans}, {Petiteau}, {Porter}, {Schutz},
  {Sesana}, {Stebbins}, {Sumner}, {Vallisneri}, {Vitale}, {Volonteri}, \&
  {Ward}}]{eLISA2}
------. 2012{\natexlab{b}}, ArXiv e-prints, 1201.3621

\bibitem[{{Amaro-Seoane} {et~al.}(2010){Amaro-Seoane}, {Sesana}, {Hoffman},
  {Benacquista}, {Eichhorn}, {Makino}, \& {Spurzem}}]{Amaro-Seoane+11}
{Amaro-Seoane}, P., {Sesana}, A., {Hoffman}, L., {Benacquista}, M., {Eichhorn},
  C., {Makino}, J., \& {Spurzem}, R. 2010, \mnras, 402, 2308, 0910.1587

\bibitem[{{Antonini} \& {Perets}(2012)}]{AP12}
{Antonini}, F., \& {Perets}, H. 2012, ArXiv e-prints, 1203.2938

\bibitem[{{Armitage} \& {Natarajan}(2005)}]{2005ApJ...634..921A}
{Armitage}, P.~J., \& {Natarajan}, P. 2005, \apj, 634, 921,
  arXiv:astro-ph/0508493

\bibitem[{{Artemova} {et~al.}(1996){Artemova}, {Bjoernsson}, \&
  {Novikov}}]{Artemova+96}
{Artemova}, I.~V., {Bjoernsson}, G., \& {Novikov}, I.~D. 1996, \apj, 461, 565

\bibitem[{Arun {et~al.}(2007{\natexlab{a}})Arun, Iyer, Sathyaprakash, \&
  Sinha}]{Arun:2007qv}
Arun, K., Iyer, B.~R., Sathyaprakash, B., \& Sinha, S. 2007{\natexlab{a}},
  Phys.Rev., D75, 124002, 0704.1086

\bibitem[{Arun {et~al.}(2007{\natexlab{b}})Arun, Iyer, Sathyaprakash, Sinha, \&
  Broeck}]{Arun:2007hu}
Arun, K., Iyer, B.~R., Sathyaprakash, B., Sinha, S., \& Broeck, C. V.~D.
  2007{\natexlab{b}}, Phys.Rev., D76, 104016, 0707.3920

\bibitem[{{Arun} {et~al.}(2009){Arun}, {Blanchet}, {Iyer}, \&
  {Sinha}}]{Arun+09}
{Arun}, K.~G., {Blanchet}, L., {Iyer}, B.~R., \& {Sinha}, S. 2009, \prd, 80,
  124018, 0908.3854

\bibitem[{{Blaes} {et~al.}(2002){Blaes}, {Lee}, \& {Socrates}}]{Bla+02}
{Blaes}, O., {Lee}, M.~H., \& {Socrates}, A. 2002, \apj, 578, 775,
  arXiv:astro-ph/0203370

\bibitem[{{Brouwer}(1959)}]{bro59}
{Brouwer}, D. 1959, \aj, 64, 378

\bibitem[{{Brown} \& {Zimmerman}(2010)}]{2010PhRvD..81b4007B}
{Brown}, D.~A., \& {Zimmerman}, P.~J. 2010, \prd, 81, 024007, 0909.0066

\bibitem[{{Buonanno} {et~al.}(2006){Buonanno}, {Chen}, \&
  {Damour}}]{Buonanno+06}
{Buonanno}, A., {Chen}, Y., \& {Damour}, T. 2006, \prd, 74, 104005,
  arXiv:gr-qc/0508067

\bibitem[{{Damour} \& {Deruelle}(1985)}]{DD85}
{Damour}, T., \& {Deruelle}, N. 1985, Journal des Astronomes Francais, 25, 21

\bibitem[{{Dirac}(1950)}]{Dirac}
{Dirac}, P.~A.~M. 1950, Can.~J.~Math., 2, 937

\bibitem[{{Dotti} {et~al.}(2012){Dotti}, {Sesana}, \& {Decarli}}]{Dotti+12}
{Dotti}, M., {Sesana}, A., \& {Decarli}, R. 2012, Advances in Astronomy, 2012,
  1111.0664

\bibitem[{{Eggleton} {et~al.}(2007){Eggleton}, {Kisseleva-Eggleton}, \&
  {Dearborn}}]{Eggleton+07}
{Eggleton}, P.~P., {Kisseleva-Eggleton}, L., \& {Dearborn}, X. 2007, in IAU
  Symposium, Vol. 240, IAU Symposium, ed. {W.~I.~Hartkopf, E.~F.~Guinan, \&
  P.~Harmanec}, 347--355

\bibitem[{{Fabrycky} \& {Tremaine}(2007)}]{Dan}
{Fabrycky}, D., \& {Tremaine}, S. 2007, \apj, 669, 1298, 0705.4285

\bibitem[{{Finn} \& {Lommen}(2010)}]{Finn+10}
{Finn}, L.~S., \& {Lommen}, A.~N. 2010, \apj, 718, 1400, 1004.3499

\bibitem[{{Ford} {et~al.}(2000{\natexlab{a}}){Ford}, {Joshi}, {Rasio}, \&
  {Zbarsky}}]{Ford00Pls}
{Ford}, E.~B., {Joshi}, K.~J., {Rasio}, F.~A., \& {Zbarsky}, B.
  2000{\natexlab{a}}, \apj, 528, 336, arXiv:astro-ph/9905347

\bibitem[{{Ford} {et~al.}(2000{\natexlab{b}}){Ford}, {Kozinsky}, \&
  {Rasio}}]{Ford00}
{Ford}, E.~B., {Kozinsky}, B., \& {Rasio}, F.~A. 2000{\natexlab{b}}, \apj, 535,
  385

\bibitem[{{Galaviz} \& {Br{\"u}gmann}(2011)}]{Galaviz+11}
{Galaviz}, P., \& {Br{\"u}gmann}, B. 2011, \prd, 83, 084013, 1012.4423

\bibitem[{{Harrington}(1968)}]{Har68}
{Harrington}, R.~S. 1968, \aj, 73, 190

\bibitem[{{Harrington}(1969)}]{Har69}
------. 1969, Celestial Mechanics, 1, 200

\bibitem[{{Hoffman} \& {Loeb}(2007)}]{Loren}
{Hoffman}, L., \& {Loeb}, A. 2007, \mnras, 377, 957, arXiv:astro-ph/0612517

\bibitem[{{Holman} {et~al.}(1997){Holman}, {Touma}, \& {Tremaine}}]{Hol+97}
{Holman}, M., {Touma}, J., \& {Tremaine}, S. 1997, \nat, 386, 254

\bibitem[{{Innanen}(1979)}]{Innanen79}
{Innanen}, K.~A. 1979, \aj, 84, 960

\bibitem[{{Innanen}(1980)}]{Innanen80}
------. 1980, \aj, 85, 81

\bibitem[{{Ivanova} {et~al.}(2010){Ivanova}, {Chaichenets}, {Fregeau},
  {Heinke}, {Lombardi}, \& {Woods}}]{Iva+10}
{Ivanova}, N., {Chaichenets}, S., {Fregeau}, J., {Heinke}, C.~O., {Lombardi},
  J.~C., \& {Woods}, T.~E. 2010, \apj, 717, 948, 1001.1767

\bibitem[{{Jaranowski} \& {Sch{\"a}fer}(1998)}]{JS98}
{Jaranowski}, P., \& {Sch{\"a}fer}, G. 1998, \prd, 57, 7274,
  arXiv:gr-qc/9712075

\bibitem[{{Jaranowski} \& {Sch{\"a}fer}(2001)}]{JS01E}
------. 2001, \prd, 63, 029902

\bibitem[{{Katz} {et~al.}(2011){Katz}, {Dong}, \& {Malhotra}}]{Boaz2}
{Katz}, B., {Dong}, S., \& {Malhotra}, R. 2011, ArXiv e-prints, 1106.3340

\bibitem[{{Kiseleva} {et~al.}(1998){Kiseleva}, {Eggleton}, \&
  {Mikkola}}]{1998KEM}
{Kiseleva}, L.~G., {Eggleton}, P.~P., \& {Mikkola}, S. 1998, MNRAS, 300, 292

\bibitem[{{Kocsis} \& {Levin}(2011)}]{2011arXiv1109.4170K}
{Kocsis}, B., \& {Levin}, J. 2011, ArXiv e-prints, 1109.4170

\bibitem[{{Kocsis} {et~al.}(2012){Kocsis}, {Ray}, \& {Portegies
  Zwart}}]{2012ApJ...752...67K}
{Kocsis}, B., {Ray}, A., \& {Portegies Zwart}, S. 2012, \apj, 752, 67,
  1110.6172

\bibitem[{{Kozai}(1962)}]{Kozai}
{Kozai}, Y. 1962, \aj, 67, 591

\bibitem[{{Krymolowski} \& {Mazeh}(1999)}]{KM99}
{Krymolowski}, Y., \& {Mazeh}, T. 1999, \mnras, 304, 720

\bibitem[{{Kulkarni} \& {Loeb}(2012)}]{Kulkarni+12}
{Kulkarni}, G., \& {Loeb}, A. 2012, \mnras, 2670, 1107.0517

\bibitem[{{Lidov}(1962)}]{Lidov}
{Lidov}, M.~L. 1962, planss, 9, 719

\bibitem[{{Lidov} \& {Ziglin}(1976)}]{Lidov+76}
{Lidov}, M.~L., \& {Ziglin}, S.~L. 1976, Celestial Mechanics, 13, 471

\bibitem[{{Lithwick} \& {Naoz}(2011)}]{LN}
{Lithwick}, Y., \& {Naoz}, S. 2011, ArXiv e-prints, 1106.3329

\bibitem[{{Lousto} \& {Nakano}(2008)}]{Lousto+08}
{Lousto}, C.~O., \& {Nakano}, H. 2008, Classical and Quantum Gravity, 25,
  195019, 0710.5542

\bibitem[{{Mandel} {et~al.}(2008){Mandel}, {Brown}, {Gair}, \&
  {Miller}}]{Ilya08}
{Mandel}, I., {Brown}, D.~A., {Gair}, J.~R., \& {Miller}, M.~C. 2008, \apj,
  681, 1431, 0705.0285

\bibitem[{{Marchal}(1990)}]{Maechal90}
{Marchal}, C. 1990, {The three-body problem}, ed. {Marchal, C.}

\bibitem[{{Mardling}(2007)}]{Mardling07}
{Mardling}, R.~A. 2007, \mnras, 382, 1768, 0706.0224

\bibitem[{{Mardling} \& {Aarseth}(2001)}]{Mardling+01}
{Mardling}, R.~A., \& {Aarseth}, S.~J. 2001, \mnras, 321, 398

\bibitem[{{Mazeh} \& {Shaham}(1979)}]{Mazeh+79}
{Mazeh}, T., \& {Shaham}, J. 1979, AA, 77, 145

\bibitem[{{McKenna} \& {Lyne}(1988)}]{ML88}
{McKenna}, J., \& {Lyne}, A.~G. 1988, \nat, 336, 226

\bibitem[{{Miller} \& {Hamilton}(2002)}]{MH02}
{Miller}, M.~C., \& {Hamilton}, D.~P. 2002, \apj, 576, 894,
  arXiv:astro-ph/0202298

\bibitem[{{Misner} {et~al.}(1973){Misner}, {Thorne}, \&
  {Wheeler}}]{Gravitation}
{Misner}, C.~W., {Thorne}, K.~S., \& {Wheeler}, J.~A. 1973, {Gravitation}, ed.
  {Misner, C.~W., Thorne, K.~S., \& Wheeler, J.~A.}

\bibitem[{{Moore}(1993)}]{Moore93}
{Moore}, C. 1993, Physical Review Letters, 70, 3675

\bibitem[{{Morais} \& {Giuppone}(2012)}]{MG12}
{Morais}, M.~H.~M., \& {Giuppone}, C.~A. 2012, ArXiv e-prints, arXiv:1204.4718

\bibitem[{{Murray} \& {Dermott}(2000)}]{MD00}
{Murray}, C.~D., \& {Dermott}, S.~F. 2000, {Solar System Dynamics}, ed.
  {Murray, C.~D.~\& Dermott, S.~F.}

\bibitem[{{Naoz} {et~al.}(2011){Naoz}, {Farr}, {Lithwick}, {Rasio}, \&
  {Teyssandier}}]{Naoz11}
{Naoz}, S., {Farr}, W.~M., {Lithwick}, Y., {Rasio}, F.~A., \& {Teyssandier}, J.
  2011, \nat, 473, 187, 1011.2501

\bibitem[{{Naoz} {et~al.}(2013){Naoz}, {Farr}, {Lithwick}, {Rasio}, \&
  {Teyssandier}}]{Naoz+11sec}
------. 2013, \mnras, 431, 2155

\bibitem[{{Naoz} {et~al.}(2012){Naoz}, {Farr}, \& {Rasio}}]{Naoz+12bin}
{Naoz}, S., {Farr}, W.~M., \& {Rasio}, F.~A. 2012, \apjl, 754, L36, 1206.3529

\bibitem[{{Nowak} \& {Wagoner}(1991)}]{NW91}
{Nowak}, M.~A., \& {Wagoner}, R.~V. 1991, \apj, 378, 656

\bibitem[{{O'Leary} {et~al.}(2009){O'Leary}, {Kocsis}, \&
  {Loeb}}]{2009MNRAS.395.2127O}
{O'Leary}, R.~M., {Kocsis}, B., \& {Loeb}, A. 2009, \mnras, 395, 2127,
  0807.2638

\bibitem[{{O'Leary} {et~al.}(2006){O'Leary}, {Rasio}, {Fregeau}, {Ivanova}, \&
  {O'Shaughnessy}}]{OLeary_06}
{O'Leary}, R.~M., {Rasio}, F.~A., {Fregeau}, J.~M., {Ivanova}, N., \&
  {O'Shaughnessy}, R. 2006, \apj, 637, 937, arXiv:astro-ph/0508224

\bibitem[{{Perets} \& {Fabrycky}(2009)}]{PF09}
{Perets}, H.~B., \& {Fabrycky}, D.~C. 2009, \apj, 697, 1048, 0901.4328

\bibitem[{{Peters}(1964)}]{Peters64}
{Peters}, P.~C. 1964, Physical Review, 136, 1224

\bibitem[{{Pribulla} \& {Rucinski}(2006)}]{Pri+06}
{Pribulla}, T., \& {Rucinski}, S.~M. 2006, \aj, 131, 2986,
  arXiv:astro-ph/0601610

\bibitem[{{Prodan} \& {Murray}(2012)}]{Prodan+12}
{Prodan}, S., \& {Murray}, N. 2012, \apj, 747, 4, 1110.6655

\bibitem[{{Sch{\"a}fer}(1987)}]{Schafer87}
{Sch{\"a}fer}, G. 1987, Physics Letters A, 123, 336

\bibitem[{{Sesana}(2010)}]{2010ApJ...719..851S}
{Sesana}, A. 2010, \apj, 719, 851, 1006.0730

\bibitem[{{Seto}(2012)}]{Seto12}
{Seto}, N. 2012, \prd, 85, 064037, 1202.4761

\bibitem[{{Shapiro} {et~al.}(1972){Shapiro}, {Pettengill}, {Ash}, {Ingalls},
  {Campbell}, \& {Dyce}}]{Shapiro+72}
{Shapiro}, I.~I., {Pettengill}, G.~H., {Ash}, M.~E., {Ingalls}, R.~P.,
  {Campbell}, D.~B., \& {Dyce}, R.~B. 1972, Physical Review Letters, 28, 1594

\bibitem[{{Sharpee} \& {Thompson}(2012)}]{Sharpee+12}
{Sharpee}, B.~J., \& {Thompson}, T.~A. 2012, ArXiv e-prints, 1204.1053

\bibitem[{{Takeda} {et~al.}(2008){Takeda}, {Kita}, \& {Rasio}}]{Takeda}
{Takeda}, G., {Kita}, R., \& {Rasio}, F.~A. 2008, \apj, 683, 1063, 0802.4088

\bibitem[{{Thompson}(2010)}]{Tho10}
{Thompson}, T.~A. 2010, ArXiv e-prints, 1011.4322

\bibitem[{{Tokovinin}(1997)}]{T97}
{Tokovinin}, A.~A. 1997, Astronomy Letters, 23, 727

\bibitem[{{Valtonen} \& {Karttunen}(2006)}]{3book}
{Valtonen}, M., \& {Karttunen}, H. 2006, {The Three-Body Problem}, ed.
  {Valtonen, M.~\& Karttunen, H.}

\bibitem[{{Valtonen}(1996)}]{Valtonen96}
{Valtonen}, M.~J. 1996, \mnras, 278, 186

\bibitem[{{Wen}(2003)}]{Wen}
{Wen}, L. 2003, \apj, 598, 419, arXiv:astro-ph/0211492

\bibitem[{{Wu} {et~al.}(2007){Wu}, {Murray}, \& {Ramsahai}}]{Wu+07}
{Wu}, Y., {Murray}, N.~W., \& {Ramsahai}, J.~M. 2007, \apj, 670, 820, 0706.0732

\bibitem[{Yunes {et~al.}(2009)Yunes, Arun, Berti, \& Will}]{Yunes:2009yz}
Yunes, N., Arun, K., Berti, E., \& Will, C.~M. 2009, Phys.Rev., D80, 084001,
  0906.0313

\bibitem[{{Yunes} {et~al.}(2011){Yunes}, {Miller}, \&
  {Thornburg}}]{Yunes:2010sm}
{Yunes}, N., {Miller}, M.~C., \& {Thornburg}, J. 2011, \prd, 83, 044030,
  1010.1721

\bibitem[{{Zhang} {et~al.}(2013){Zhang}, {Hamilton}, \& {Matsumura}}]{Zhang+13}
{Zhang}, K., {Hamilton}, D.~P., \& {Matsumura}, S. 2013, ArXiv e-prints,
  1302.1620

\end{thebibliography}

\end{document}